%% file: mued.tex
\documentclass[12pt]{iopart}

\usepackage{epsf}
\usepackage{graphicx}
\usepackage{color}
\usepackage{axodraw}
\usepackage{cite}

\newcommand\be{\begin{eqnarray}}
\newcommand\ee{\end{eqnarray}}

\newcommand\ben{\begin{eqnarray*}}
\newcommand\een{\end{eqnarray*}}
\newcommand\bc{\begin{center}}
\newcommand\ec{\end{center}}
\newcommand\g{\gamma}

\newcommand\intez{\int_{0}^{\pi R} d y}
\newcommand\summ{\sum_{n=1}^{\infty}}

\newcommand\cc{\g^\mu \frac{1}{\sqrt 2} \frac{1}{16\pi^2}\ln \left ( \frac{\Lambda}{\mu}\right )^2}

\newcommand{\beq}{\begin{equation}}
\def\eeq#1{\label{#1}\end{equation}}
\def\beqa{\begin{eqnarray}}
\def\eeqa#1{\label{#1}\end{eqnarray}}

\begin{document} 

\begin{flushright}
HRI-P-10-02-001 \\
RECAPP-HRI-2010-006 
\end{flushright}

\title[]{Minimal Universal Extra Dimensions in {\tt CalcHEP}/{\tt CompHEP}}

\author{AseshKrishna Datta}
\address{Regional Centre for Accelerator-based Particle Physics (RECAPP) \\
Harish-Chandra Research Institute, Chhatnag Road, Jhusi, Allahabad - 211 019, India}
\ead{asesh@hri.res.in}

\author{Kyoungchul Kong}
\address{Theoretical Physics Department, SLAC, Menlo Park, CA 94025, USA} 
\ead{kckong@slac.stanford.edu}

\author{Konstantin T. Matchev}
\address{Physics Department, University of Florida, Gainesville, FL 32611, USA} 
\ead{matchev@phys.ufl.edu}

\begin{abstract}
We present an implementation of the model of minimal universal extra dimensions 
(MUED) in {\tt CalcHEP}/{\tt CompHEP}. We include all level-1 and level-2 Kaluza-Klein 
(KK) particles outside the Higgs sector. The mass spectrum is automatically 
calculated at one loop in terms of the two input parameters in MUED: 
the inverse radius $R^{-1}$ of the extra dimension and the 
cut-off scale of the model $\Lambda$. We implement both the KK number conserving 
and the KK number violating interactions of the KK particles.
We also account for the proper running of the gauge coupling constants above
the electroweak scale.
The implementation has been extensively cross-checked against known analytical 
results in the literature and numerical results from other programs. 
Our files are publicly available and can be used to perform various 
automated calculations within the MUED model.
\end{abstract}


\maketitle

\section{Introduction}
\label{sec:intro}

The Standard Model (SM) of particle physics has been successfully verified
by experiment at low energies.  Nevertheless, even if the Higgs boson is discovered,
the SM will still be considered to be an incomplete theory, as it fails to
provide the long-sought missing link between Einstein's General Relativity 
and Quantum Mechanics.
The leading candidate for a quantum theory of gravity, string theory,
typically posits the existence of several new ingredients, which are
absent in the SM: new spatial dimensions, a symmetry between bosons 
and fermions (supersymmetry), as well as new gauge interactions.
All of these new ingredients are manifestly present at the Planck scale, 
but it is not at all clear which of them survive down to low energies.
Traditionally, supersymmetry and extra gauge interactions have attracted 
the most attention, and their consequences for collider phenomenology 
have been extensively studied \cite{Chung:2003fi,Langacker:2008yv}.
Within the last 10 years or so, there has been a resurgence of interest in 
models with extra spatial dimensions, whose presence might be revealed
in high energy collider experiments such as the Tevatron at Fermilab, 
the Large Hadron Collider (LHC) at CERN, or the proposed International 
Linear Collider (ILC). By now a whole plethora of extra-dimensional models
have been described and studied to various extent in the literature.
Roughly speaking, they can all be classified according to the following two
criteria:
\begin{itemize}
\item How many and which of the SM particles can access the extra dimensions (the bulk).
The two extremes here are provided by the ``large'' extra dimension models
(also known as ADD, after the initials of their original proponents)
\cite{ArkaniHamed:1998rs}, in which only gravity can enter into the bulk,
and the Universal Extra Dimensions (UED) models \cite{Appelquist:2000nn}, 
in which {\em all} SM particles are allowed to propagate in the bulk.
\item What is the metric of the bulk. It can be flat (e.g.~in UED), 
or warped \cite{Randall:1999ee}.
\end{itemize}

In this paper, we shall concentrate on the simplest case of a single
flat extra dimension, which is accessible to the 
full SM particle content \cite{Appelquist:2000nn} 
(see Refs. \cite{Dobrescu:2004zi,Burdman:2006gy,Dobrescu:2007xf,Dobrescu:2007ec,Freitas:2007rh,Ghosh:2008ix} 
for the case of two universal extra dimensions). 
This particular scenario has recently been studied in relation to 
collider phenomenology 
\cite{Rizzo:2001sd,Macesanu:2002db,Cheng:2002ab,Carone:2003ms,
Bhattacharyya:2005vm,Battaglia:2005zf,Smillie:2005ar,Battaglia:2005ma,Datta:2005zs,
Datta:2005vx,Kong:2006pi,Cembranos:2006gt,Bhattacherjee:2007wy,Bhattacherjee:2008ik,
Konar:2009ae,Matsumoto:2009tb,Bhattacharyya:2009br,Bandyopadhyay:2009gd}, 
indirect low-energy constraints 
\cite{Agashe:2001ra,Agashe:2001xt,Appelquist:2001jz,%
Petriello:2002uu,Appelquist:2002wb,Chakraverty:2002qk,Buras:2002ej,%
Oliver:2002up,Buras:2003mk,Iltan:2003tn,Khalil:2004qk,Bashiry:2008en,Gogoladze:2006br,Haisch:2007vb},
dark matter 
\cite{Servant:2002aq,Cheng:2002ej,Servant:2002hb,Majumdar:2002mw,Burnell:2005hm,Kong:2005hn,Arrenberg:2008wy,
Majumdar:2003dj,Kakizaki:2005en,Kakizaki:2005uy,Bertone:2002ms,Bergstrom:2004cy,
Baltz:2004ie,Bergstrom:2004nr,Bringmann:2005pp,Barrau:2005au,Birkedal:2005ep,Flacke:2008ne,Belanger:2008gy,
Blennow:2009ag,Matsumoto:2007dp,Kakizaki:2006dz,Matsumoto:2005uh}
and
cosmology 
\cite{Matsumoto:2006bf,Shah:2006gs,Li:2005aia,Feng:2003nr,Mazumdar:2003vg,Bringmann:2003sz,Kolb:2003mm}.
It is therefore of great interest to have an implementation 
of the Minimal UED model (reviewed below in Section~\ref{sec:MUED})
in the most popular general purpose computer programs for 
collider and astroparticle phenomenology. The main goal
of this paper is to present one such implementation, suitable for
either {\tt CalcHEP} \cite{Pukhov:2004ca} or {\tt CompHEP} \cite{Pukhov:1999gg}.
There are several advantages of choosing {\tt CalcHEP} and {\tt CompHEP} for this purpose:
\begin{itemize}
\item {\tt CalcHEP} and {\tt CompHEP} can be used for
parton-level event generation, preserving the 
full spin correlations in both production and decay.
\item {\tt CalcHEP} and {\tt CompHEP} can be easily interfaced \cite{Belyaev:2000wn}
to a general purpose event generator such as {\tt PYTHIA}
\cite{Sjostrand:2003wg} for the simulation of fragmentation, hadronization
and showering.
\item {\tt CalcHEP} and {\tt CompHEP} can be easily interfaced
with a dark matter program such as {\tt micrOMEGAs}
\cite{Belanger:2006is} for the calculation of the 
relic density and detection rates of a generic dark matter
candidate.
\item The implementation of new models is very straightforward 
and user-friendly, as we shall demonstrate below with the example
of Minimal UED.
\end{itemize}

The paper is organized as follows. In Section~\ref{sec:MUED}
we first review the Minimal UED model (MUED), introducing the 
relevant new particles, couplings and interactions. 
In Section~\ref{sec:chep} we explain how those were incorporated 
in {\tt CalcHEP} and {\tt CompHEP}. Throughout the paper we 
assume that the readers are already familiar with these programs, 
so that we only need to explain the additional *.mdl model files 
related to our UED implementation\footnote{Our implementation was
originally developed for the Second MC4BSM workshop in Princeton, 
March 24-27, 2007.
Since then, the Minimal UED model has been partially implemented 
in {\tt PYTHIA} \cite{ElKacimi:2009zj}, and more fully in {\tt CalcHEP}, 
{\tt MadGraph}, {\tt PYTHIA} or {\tt Sherpa} through 
{\tt FeynRules} \cite{Christensen:2008py,Christensen:2009jx}.}.
In Section~\ref{sec:pheno} we discuss how the implementation 
can be used to study the collider phenomenology of MUED
and show some illustrative results.
In the Appendices we list some more technical results 
which may be useful to some readers. For example, 
\ref{app:5d} contains the five-dimensional UED Lagrangian
and
\ref{app:vertices} contains the resulting Feynman 
rules for the level 1 KK particles after compactification.

\section{The Minimal UED Model}
\label{sec:MUED}

\subsection{KK decomposition}
\label{sec:dec}

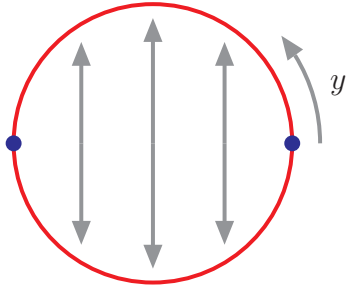
\begin{figure}[tb]
\unitlength=1.5 pt
\SetScale{1.5}
\SetWidth{1.0}      
\normalsize    
{} \qquad\allowbreak
\centerline{
\begin{picture}(140,110)(20,30)
\SetColor{Gray}
\LongArrow( 75, 75)( 75, 45)
\LongArrow( 75, 75)( 75,105)
\LongArrow( 57, 75)( 57, 51)
\LongArrow( 57, 75)( 57, 99)
\LongArrow( 93, 75)( 93, 51)
\LongArrow( 93, 75)( 93, 99)
\LongArrowArc(75,75)(42,0,40)
\Text(122,90)[c]{\Black{$y$}}
\SetColor{Red}
\CArc( 75, 75)(35,0,360)
\SetColor{Blue}
\Vertex( 40, 75){2}
\Vertex(110, 75){2}
\end{picture} 
}
\caption{The $S_1/Z_2$ compactification of a single extra dimension
on a circle with opposite points identified, as
indicated by the grey arrows.
The blue dots represent the fixed (boundary) points
and $y$ is the coordinate along the extra dimension.
}
\label{fig:UED} 
\end{figure}

The five-dimensional (5D) UED model \cite{Appelquist:2000nn}
is simply the Standard Model placed in 
an extra dimension compactified on an $S_1/Z_2$ orbifold, 
as shown in Fig.~\ref{fig:UED}. Let us label the usual $3+1$ 
space-time dimensions with $x^\mu$, $\mu=0,1,2,3$, reserving 
the coordinate $y$ for the extra dimension. 
In order to end up with chiral fermions in 4 dimensions
and to project out unwanted gauge degrees of freedom, 
one typically imposes an additional symmetry, thus
creating a manifold with boundaries. For example,
in the case of the $S_1/Z_2$ orbifold 
shown in Fig.~\ref{fig:UED}, one identifies 
the opposite points on the circle, which
creates two fixed points, denoted with the blue dots.
Any 5-dimensional field can now be assigned a definite parity 
with respect to the orbifold projection
${\cal P}_5 : y \rightarrow -y$.
For example, consider a generic scalar field $\phi (x,y)$.
An even scalar field $\phi^+ (x,y)$ is expanded in Kaluza-Klein (KK) modes as
\begin{equation}
\phi^+ (x,y) = \frac{1}{\sqrt{\pi R}}\, \phi^+_0(x) 
+ \frac{2}{\sqrt{\pi R}} \, \summ \phi^+_n (x) \cos\frac{ny}{R} \, ,
\end{equation}
and obeys Neumann boundary conditions at the two fixed points:
\begin{equation}
 \left( \frac{\partial \phi^+ (x,y)}{\partial y} \right)_{y=0} 
=\left( \frac{\partial \phi^+ (x,y)}{\partial y} \right)_{y=\pi R} 
= 0.
\label{NeuBC}
\end{equation}
Here $x$ is the usual 4-dimensional spacetime coordinate $x^\mu$, 
$R$ is the size of the extra dimension and $n$ labels the KK-level.
The SM modes correspond to $n=0$. In contrast,
the KK decomposition of an odd scalar field 
\begin{equation}
\phi^- (x,y) = \frac{2}{\sqrt{\pi R}} \, \summ \phi^-_n (x) \sin\frac{ny}{R} \, ,
\end{equation}
is missing a zero mode ($n=0$) and obeys Dirichlet boundary conditions 
\begin{equation}
\phi^- (x,0) = \phi^- (x,\pi R) = 0.
\label{DirBC}
\end{equation}

One can similarly assign a definite ${\cal P}_5$ parity to each
component of a gauge field $A_M (x,y)$, $M=0,1,2,3,5$. The usual $3+1$ 
components $A_\mu$, $\mu=0,1,2,3$, are chosen to be even, 
which ensures the presence of the SM gauge fields $A_\mu^{0}(x)$
at the $n=0$ level, while the extra-dimensional component 
$A_5$ is taken to be odd. The corresponding 
KK expansions of the 5-dimensional gauge fields are given by
\be
A_\mu(x,y) &=& \frac{1}{\sqrt{\pi R}} \left \{ A_\mu^{0}(x) + \sqrt{2} \summ A_\mu^{n} (x) \cos (\frac{ny}{R}) \right  \} \, ,\\[2mm]
A_5  (x,y) &=& \sqrt{\frac{2}{\pi R}} \summ A_5^n (x) \sin (\frac{n y}{R}) \, .
\ee
At the two fixed points $y=0$ and $y=\pi R$, the components $A_\mu(x,y)$ ($A_5(x,y)$) 
obey Neumann (Dirichlet) boundary conditions analogous to
eq.~(\ref{NeuBC}) (eq.~(\ref{DirBC})).

The KK decomposition of a fermion is rather interesting.
Since there is no chirality in 5 dimensions, the KK modes of the SM
fermions come in vectorlike pairs, i.e. there is a left-handed 
{\em and} a right-handed KK mode for each SM chiral fermion. 
For example, the $SU(2)_W$-singlet chiral fermions $\psi_R^0 (x)$ of the SM 
(which happen to be all right-handed) are 
obtained from the following decomposition
\be
\psi^+_R (x,y) &=& \frac{1}{\sqrt{2\pi R}}\psi_R^0 (x) + 
                 \frac{1}{\sqrt{\pi R}}\summ \psi_R^n (x) \cos\frac{n y}{R} \, , 
\label{psiR+}\\ [2mm] 
\psi^-_R (x,y) &=&  
                 \frac{1}{\sqrt{\pi R}}\summ \psi_L^n (x) \sin\frac{n y}{R} \, ,
\label{psiR-}
\ee
where upon compactification, the two KK fermions $\psi_R^n (x)$ and $\psi_L^n (x)$
at any given KK level $n$ pair up to give a Dirac fermion of mass $\frac{n}{R}$.
Similarly, the $SU(2)_W$-doublet SM fermions $\Psi_L^0 (x)$ (which happen to be left-handed)
arise from
\be
\Psi^+_L (x,y) &=& \frac{1}{\sqrt{2\pi R}}\Psi_L^0 (x) + 
                 \frac{1}{\sqrt{\pi R}}\summ \Psi_L^n (x) \cos\frac{n y}{R} \, ,
\label{psiL+}\\ [2mm] 
\Psi^-_L (x,y) &=&  
                 \frac{1}{\sqrt{\pi R}}\summ \Psi_R^n (x) \sin\frac{n y}{R} \, ,
\label{psiL-}
\ee
where the massive Dirac fermion at each $n$ is now formed from 
$\Psi_L^n (x)$ and $\Psi_R^n (x)$.

From eqs.~(\ref{psiR+}-\ref{psiL-}) we see that there exist 
{\em left-handed} KK modes $\psi_L^n (x)$, which are associated
with the {\em right-handed} SM fermions $\psi_R^0 (x)$ and vice versa --- 
there are {\em right-handed} KK modes $\Psi_R^n (x)$, which go along with
the {\em left-handed} SM fermions $\Psi_L^0 (x)$.
This often leads to some confusion in the literature
when it comes to the labelling of fermion KK partners.
It should be understood that the chiral index ($L$ or $R$) 
of a KK mode fermion refers to the chirality of its SM partner.
Here we shall also utilize an alternative convention,
introduced in \cite{Cheng:2002iz}, where the KK fermions are
identified by their $SU(2)_W$ quantum numbers instead: 
$SU(2)_W$-doublets ($SU(2)_W$-singlets) are denoted 
with capital (lowercase) letters. This convention was already 
employed in eqs.~(\ref{psiR+}-\ref{psiL-}) as well.
With those conventions, the fermion content of the Minimal
UED model is listed in Table~\ref{tab:fermion}.
\begin{table}[t]
\caption{Fermion content of the Minimal UED model.
$SU(2)_W$-doublets ($SU(2)_W$-singlets) are denoted 
with capital (lowercase) letters. KK modes carry 
a KK index $n$, and for simplicity we omit the 
index ``0'' for the SM zero modes. 
\label{tab:fermion}}
\footnotesize\rm
\begin{tabular}{cccc}
\hline \hline
$SU(2)_W$ representations	& SM mode	& KK modes	\\
\hline \hline
Quark doublet &
$q_L (x) = \left ( \begin{array}{c} U_L (x) \\ D_L (x)   \end{array} \right )$ &
$Q^n_L (x) = \left ( \begin{array}{c} U^n_L (x) \\ D^n_L (x)   \end{array}\right )$,
$Q^n_R (x) = \left ( \begin{array}{c} U^n_R (x) \\ D^n_R (x)   \end{array}\right )$ \\
\hline
Lepton doublet &
$L_L (x) = \left ( \begin{array}{c} \nu_L (x) \\ E_L (x)   \end{array} \right )$ &
$L^n_L (x) = \left ( \begin{array}{c} \nu^n_L (x) \\ E^n_L (x)   \end{array}\right )$,
$L^n_R (x) = \left ( \begin{array}{c} \nu^n_R (x) \\ E^n_R (x)   \end{array}\right )$ \\
\hline
Quark Singlet &
$u_R(x)$ &
$u^n_R (x)$, $u^n_L (x)$ \\
\hline
Quark Singlet &
$d_R(x)$ &
$d^n_R (x)$, $d^n_L (x)$ \\
\hline
Lepton Singlet &
$e_R(x)$ &
$e^n_R (x)$, $e^n_L (x)$ \\
\hline
\end{tabular}
\end{table}

Finally, notice that the geometry in Fig.~\ref{fig:UED} is still invariant 
under the interchange of the two fixed points.  
The corresponding $Z_2$ symmetry is the celebrated KK parity
and will be a symmetry of the Lagrangian as long as one continues
to treat the two boundary points in a symmetric fashion.

\subsection{KK mass spectrum}

\begin{figure}[tb]
\centerline{
\includegraphics[width=.47\linewidth]{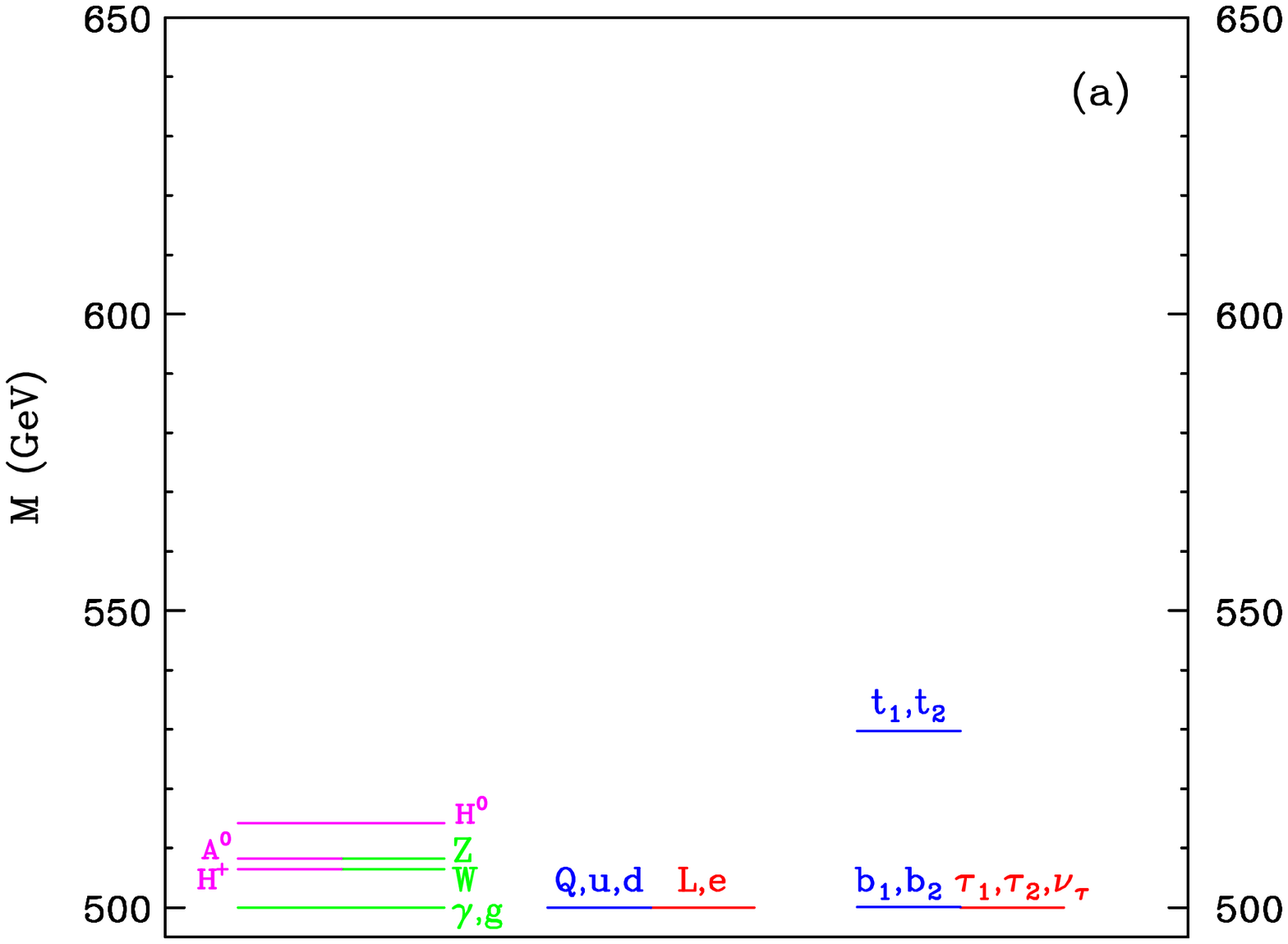} \hspace{0.1cm}
\includegraphics[width=.47\linewidth]{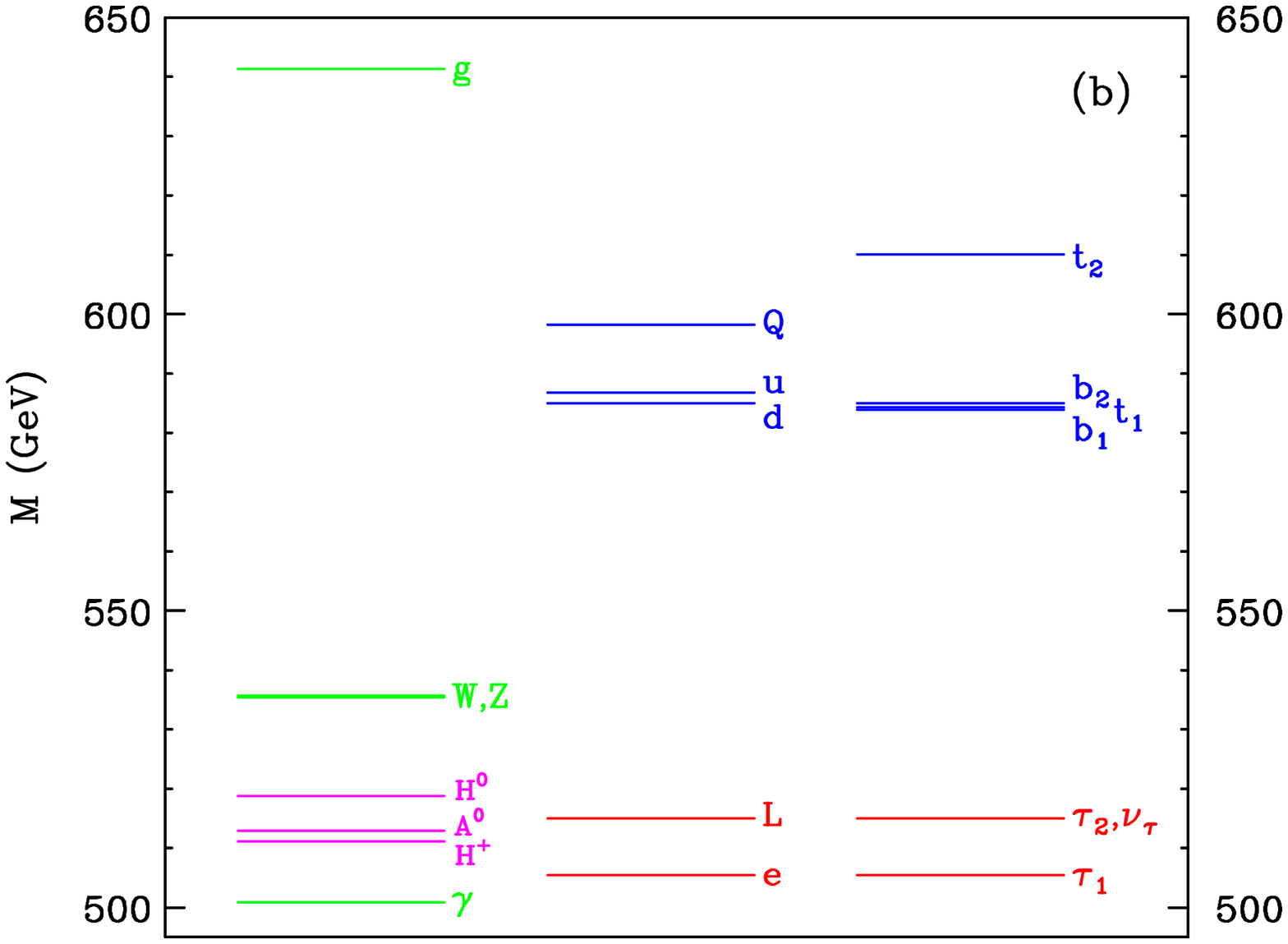}
}
\caption{\label{fig:spectrum} The spectrum of the first KK level
at (a) tree level and (b) one-loop, for 
$R^{-1}=500$ GeV, $\Lambda R = 20$, $m_h=120$ GeV,
and assuming vanishing boundary terms at the cut-off scale $\Lambda$.
(From Ref.~\cite{Cheng:2002iz}.)}
\end{figure}

At tree level, the mass $m_n$ of any KK mode at the $n$-th KK level is given by
\be
m_n^2 = \frac{n^2}{R^2} + m_0^2 \, ,
\ee
where $R$ is the radius of the extra dimension as illustrated in Fig.~\ref{fig:UED},
and $m_0$ is the mass of the corresponding SM particle (zero mode).
The resulting mass spectrum for the first KK level is shown in Fig.~\ref{fig:spectrum}a 
for $R^{-1}=500$ GeV, and can be seen to be highly degenerate.
In fact, several of the lightest $n=1$ KK modes have no allowed
decays and are absolutely stable. 

However, this drastic conclusion is completely reversed, once
radiative corrections are taken into account \cite{Cheng:2002iz}.
First, the mass spectrum gets renormalized by bulk interactions, 
which are uniquely fixed in terms of the SM gauge and Yukawa couplings, 
and thus contain no new parameters beyond those already appearing in the SM.
At the same time, the KK masses also receive contributions from
terms localized on the boundary points (the two blue dots in Fig.~\ref{fig:UED}).
The coefficients of the boundary terms are in principle 
new free parameters of the theory. The Minimal UED model 
makes the ansatz that all boundary terms simultaneously 
vanish at some high scale $\Lambda > R^{-1}$. The boundary
terms are then regenerated at lower scales through RGE 
running, and lead to additional corrections to the KK mass 
spectrum \cite{Cheng:2002iz}. The resulting one-loop 
corrected mass spectrum is shown in Fig.~\ref{fig:spectrum}b. 
The mass splittings among the different $n=1$
KK modes are now sufficiently large to allow prompt cascade decays 
to the lightest KK particle (LKP). For the parameter values
shown in the figure, the LKP turns out to be\footnote{Strictly 
speaking, the true LKP in Fig.~\ref{fig:spectrum}b is the 
KK graviton ${\cal G}_1$ (not shown). However, due to its
extremely weak couplings, ${\cal G}_1$ is irrelevant
for collider phenomenology. For its astrophysical
implications, see \cite{Feng:2003uy}.} the KK ``photon'' $\gamma_1$,
although at larger $m_h$ the LKP can also be the charged 
KK Higgs boson $H^\pm_1$ \cite{Cembranos:2006gt}.

The mass eigenstates of the KK photon $\gamma_n$ and the KK $Z$-boson $Z_n$ 
are mixtures of the corresponding interaction eigenstates: 
the KK mode $B_n$ of the hypercharge gauge boson
and the KK mode $W^3_n$ of the neutral $SU(2)_W$ gauge boson. 
The mixing angle $\theta_n$ is obtained by 
diagonalizing the mass matrix in the $(B_n,W^3_n)$ basis
\be
\left (
\begin{array}{cc}
\frac{n^2}{R^2} + \frac{1}{4} g_1^2 v^2+\hat\delta m^2_{B_n}	
&	\frac{1}{4} g_1 g_2 v^2 \\
\frac{1}{4} g_1 g_2 v^2						
& 	\frac{n^2}{R^2} + \frac{1}{4} g_2^2 v^2 +\hat\delta m^2_{W^3_n}
\end{array}
\right ) \, ,
\label{NGBmass}
\ee
where $g_1$ ($g_2$) is the hypercharge (weak) gauge coupling,
$v=246$ GeV is the vev of the SM Higgs boson,
and $\hat\delta$ represents the total one-loop correction, 
including both bulk ($\delta$) and boundary ($\bar{\delta}$) contributions \cite{Cheng:2002iz}:
\begin{equation}
\hat\delta m_{V_n}^2 \equiv \delta m_{V_n}^2  + \bar{\delta} m_{V_n}^2  \, .
\label{deltas}
\end{equation}
Note that for $n\ge 1$ the KK mixing angle $\theta_n$ is in general different 
from the zero-mode (Weinberg) angle $\theta_0\equiv \theta_W$ in the SM. 
For typical values of $R^{-1}$ and $\Lambda$, $\theta_n \ll \theta_W$,
and the neutral gauge boson KK mass eigenstates become approximately 
aligned with the corresponding interaction eigenstates:
$\gamma_n\approx B_n$ and $Z_n\approx W^3_n$ for $n\ge 1$.
This approximation will be used in our MUED implementation described below
in Section~\ref{sec:chep}.

\subsection{KK interactions}

The bulk interactions of the KK modes are already fixed by the SM.
The 5D MUED Lagrangian is a straightforward generalization of the SM 
Lagrangian to 5 dimensions, as discussed in \ref{app:5d}.
Upon compactification, integrating over the extra-dimensional coordinate
$y$, one recovers the bulk interactions among the various KK modes 
and their SM counterparts (see \ref{app:vertices}).
Since translational invariance holds in the bulk, all these bulk 
interactions conserve both KK number and KK parity. 
%
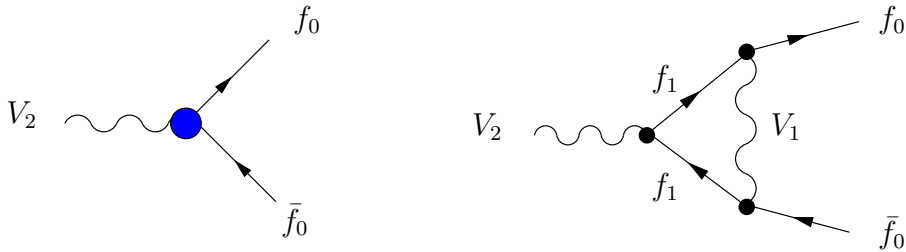
\begin{figure}[tb]
\centerline{\input{200.pstex_t}}
\caption{The effective $\bar{f}_0 V^\mu_2 f_0$ KK-number violating coupling 
on the left is generated at one loop order from the one loop diagram on the right. \label{fig:200}}
\end{figure}
%

However, as already alluded to in the previous subsection, there may also exist
``boundary'' interactions localized on the fixed points in Fig.~\ref{fig:UED}.
They do not respect translational invariance and therefore break KK number by
even units. Such interactions may already appear at the scale $\Lambda$, 
being generated by the new physics which is the ultraviolet completion of UED.
In the Minimal UED version, one makes the assumption that no such terms are present 
at the scale $\Lambda$. Even so, upon renormalization to lower energy scales,
boundary terms are radiatively generated from bulk interactions. This is illustrated 
in Fig.~\ref{fig:200}, where we show how an effective coupling between a
level 2 KK gauge boson $V_2$ and two SM fermions is generated at one loop from a diagram 
with level 1 KK particles running in the loop. This effective coupling
\ben
-i \frac{g}{\sqrt2} \left ( \frac{\bar{\delta}m^2_{A_2}}{m^2_2} - 2\frac{\bar{\delta}m_{f_2}}{m_2} \right ) 
\bar{\psi}_0 \g^\mu T^a P_+ \psi_0 A_{2\mu} 
\een
can be expressed in terms of the boundary contributions $\bar{\delta}m_n$
(see eq.~(\ref{deltas})) to the one-loop mass corrections~\cite{Cheng:2002iz}.
The explicit form of this effective coupling is summarized in 
Table~\ref{tab:200} for each different type of level 2 KK gauge boson and 
for the various possible SM fermion pairs.
\begin{table}[tb]
\caption{Boundary interactions involving level 2 KK gauge bosons and two SM fermions.
Here $I_3$ is the fermion isospin and $Y_L$ ($Y_R$) is the hypercharge of
a left-handed (right-handed) SM fermion. In the case of top quarks, one
has to include in $\bar{\delta}(m_{f_2})$ the additional corrections 
proportional to the top Yukawa coupling $h_t$:
$\bar{\delta}_{h_t}m_{T_{n}}$ and $\bar{\delta}_{h_t}m_{t_n}$, respectively 
(see~\cite{Cheng:2002iz} for details).
\label{tab:200}}
\begin{indented}
\item[]
\centerline{
\begin{tabular}{ccl}
\hline \hline
$n=2$ KK boson		& $n=0$ SM fermion	& Vertex		\\
\hline \hline
$U(1)_Y$ gauge boson	& Lepton 		& $ig_1 \cc \left [ \frac{Y_L}{2} P_L \left ( \frac{31}{24}g_1^2+\frac{27}{8}g_2^2 \right ) \right. $ \\
                        &                       & \hspace*{3cm} $\left. + \frac{Y_R}{2} P_R \left (\frac{14}{3}g_1^2 \right ) \right ]$ \\
	$B_2$		& Quark (up)		& $ig_1 \cc \left [ \frac{Y_L}{2} P_L \left ( \frac{7}{24}g_1^2+\frac{27}{8}g_2^2+6g_3^2 \right ) \right.$ \\
                        &                       & \hspace*{3cm} $\left.+ \frac{Y_R}{2} P_R \left (\frac{13}{6}g_1^2 +6g_3^2 \right ) \right ]$ \\
			& Quark (down)		& $ig_1 \cc \left [ \frac{Y_L}{2} P_L \left ( \frac{7}{24}g_1^2+\frac{27}{8}g_2^2+6g_3^2 \right ) \right.$ \\
                        &                       & \hspace*{3cm} $\left.+ \frac{Y_R}{2} P_R \left (\frac{2}{3}g_1^2 +6g_3^2 \right ) \right ]$ \\
\hline \hline
$SU(2)_W$ gauge boson	& Lepton 		& $i I_3 g_2 \cc P_L \left [ \frac{9}{8}g_1^2 - \frac{33}{8}g_2^2 \right ]$ \\
	$Z_2$		& Quark			& $i I_3 g_2 \cc P_L \left [ \frac{1}{8}g_1^2 - \frac{33}{8}g_2^2 + 6g_3^2 \right ]$ \\
\hline \hline
$SU(2)_W$ gauge boson	& Lepton 		& $i \frac{g_2}{\sqrt{2}} \cc P_L \left [ \frac{9}{8}g_1^2 - \frac{33}{8}g_2^2 \right ]$ \\
	$W_2$		& Quark			& $i \frac{g_2}{\sqrt{2}} \cc P_L \left [ \frac{1}{8}g_1^2 - \frac{33}{8}g_2^2 + 6g_3^2 \right ]$ \\
\hline \hline
$SU(3)_c$ gauge boson	& Quark (up) 		& $i g_3 \frac{\lambda^A}{2}\cc \left [ P_L \left ( \frac{1}{8}g_1^2 + \frac{27}{8}g_2^2 -\frac{11}{2}g_3^2 \right ) \right. $ \\
 & & \hspace{4cm}$\left. +P_R \left ( 2g_1^2-\frac{11}{2}g_3^2 \right ) \right ] $ \\
	$G_2$		& Quark	(down)		& $i g_3 \frac{\lambda^A}{2}\cc \left [ P_L \left ( \frac{1}{8}g_1^2 + \frac{27}{8}g_2^2 -\frac{11}{2}g_3^2 \right ) \right.$ \\
& & \hspace{4cm}$\left. +P_R \left ( \frac{1}{2}g_1^2-\frac{11}{2}g_3^2 \right ) \right ]$ \\
\hline
\end{tabular}
}
\end{indented}
\end{table}

\section{Model files}
\label{sec:chep}

Having reviewed the MUED model, we are now in a position to 
describe its implementation in {\tt CalcHEP} and {\tt CompHEP}.
Each one of these programs gives its users an opportunity to incorporate
new physics in the already existing framework of the SM, MSSM, etc.
To this end, one must simply supply an updated version of the 
four model files defining a given physics scenario in {\tt CalcHEP} and {\tt CompHEP}:
{\tt prtclsN.mdl}, {\tt varsN.mdl}, {\tt funcN.mdl} and
{\tt lgrngN.mdl}, where {\tt N} stands for the numerical label of the
physics scenario in the model menu of {\tt CalcHEP} and {\tt CompHEP}.
We shall now discuss each one of those files, which are available 
from {\tt http://home.fnal.gov/$\sim$kckong/mued/}.

\subsection{Particles\label{sec:particles}}

\begin{table}[tb]
\caption{KK gauge bosons. \label{tab:gauge}}
\begin{indented}
\item[]
\begin{tabular}{ccccccc}
\hline \hline
Name  &A  &A+ &2*spin& mass &width  &color \\
\hline \hline
$G_\mu^1$    & KG      &KG       &2     &MKG   &wKG   &8\\
$B_\mu^1$    & B1      &B1       &2     &MB1   &0     &1\\
$Z_\mu^1$    & Z1      &Z1       &2     &MZ1   &wZ1   &1\\
$W_\mu^1$    &$\sim W+$&$\sim W-$&2     &MW1   &wW1   &1\\
$G_\mu^2$    &$\sim G2$&$\sim G2$&2     &MKG2  &wKG2  &8\\
$B_\mu^2$    &B2       &B2       &2     &MB2   &wB2   &1\\
$Z_\mu^2$    &Z2       &Z2       &2     &MZ2   &wZ2   &1\\
$W_\mu^2$    &$\sim W2$&$\sim w2$&2     &MW2   &wW2   &1\\
\hline
\end{tabular}
\end{indented}
\end{table}

\begin{table}[htb]
\caption{KK leptons. \label{tab:leptons}}
\begin{indented}
\item[]
\begin{tabular}{ccccccc }
\hline \hline
Name  &A  &A+ &2*spin& mass &width &color \\
\hline \hline
$e_L^1$      &$\sim eL$&$\sim EL$&1     &DMe   &wDe1  &1 \\
$\mu_L^1$    &$\sim mL$&$\sim ML$&1     &DMm   &wDe2  &1 \\
$\tau_L^1$   &$\sim tL$&$\sim TL$&1     &DMt   &wDe3  &1 \\
$e_R^1$      &$\sim eR$&$\sim ER$&1     &SMe   &wSe1  &1 \\
$\mu_R^1$    &$\sim mR$&$\sim MR$&1     &SMm   &wSe2  &1 \\
$\tau_R^1$   &$\sim tR$&$\sim TR$&1     &SMt   &wSe3  &1 \\
$\nu_e^1$    &$\sim n1$&$\sim N1$&1     &DMen  &wDn1  &1 \\
$\nu_\mu^1$  &$\sim n2$&$\sim N2$&1     &DMmn  &wDn2  &1 \\
$\nu_\tau^1$ &$\sim n3$&$\sim N3$&1     &DMtn  &wDn3  &1 \\
\hline
$e_L^2$      &$\sim le$&$\sim lE$&1     &DMe2  &wDe12 &1\\
$\mu_L^2$    &$\sim lm$&$\sim lM$&1     &DMm2  &wDe22 &1\\
$\tau_L^2$   &$\sim lt$&$\sim lT$&1     &DMt2  &wDe32 &1\\
$e_R^2$      &$\sim re$&$\sim rE$&1     &SMe2  &wSe12 &1\\
$\mu_R^2$    &$\sim rm$&$\sim rM$&1     &SMm2  &wSe22 &1\\
$\tau_R^2$   &$\sim rt$&$\sim rT$&1     &SMt2  &wSe32 &1\\
$\nu_e^2$    &$\sim en$&$\sim eN$&1     &DMen2 &wDn12 &1\\
$\nu_\mu^2$  &$\sim mn$&$\sim mN$&1     &DMmn2 &wDn22 &1\\
$\nu_\tau^2$ &$\sim tn$&$\sim tN$&1     &DMtn2 &wDn32 &1\\
\hline
\end{tabular}
\end{indented}
\end{table}

New particles are defined in the {\tt prtclsN.mdl} model file.
We incorporate the $n=1$ and $n=2$ KK modes of 
the gauge bosons (see Table~\ref{tab:gauge}),
leptons (see Table~\ref{tab:leptons})
and quarks (see Table~\ref{tab:quarks}).
In Tables~\ref{tab:gauge}-\ref{tab:quarks}
the KK number is represented by a superscript $n=1$ or $n=2$, while
the subscript is either the Lorentz index ($\mu$) of the vector particles 
in Table~\ref{tab:gauge} or the chirality index of the fermion particles
in Tables~\ref{tab:leptons} and \ref{tab:quarks}. 
We remind the reader that all KK fermions are vectorlike and 
the chirality index refers to the chirality of their SM counterparts.
The corresponding masses and widths of the KK fermions 
in Tables~\ref{tab:leptons} and \ref{tab:quarks} carry ``D'' or ``S'' to indicate 
their nature, $SU(2)_W$-doublet or $SU(2)_W$-singlet, respectively.
The new particles listed in Tables~\ref{tab:gauge}-\ref{tab:quarks}
are in addition to the usual SM particles which are not shown here.

\begin{table}[tb]
\caption{KK quarks. \label{tab:quarks}}
\begin{indented}
\item[]
\begin{tabular}{ccccccc}
\hline \hline
Name  &A  &A+ &2*spin& mass &width &color \\
\hline \hline
$u_L^1$   &Du &DU &1     &DMu   &wDu   &3 \\
$d_L^1$   &Dd &DD &1     &DMd   &wDd   &3 \\
$c_L^1$   &Dc &DC &1     &DMc   &wDc   &3 \\
$s_L^1$   &Ds &DS &1     &DMs   &wDs   &3 \\
$t_L^1$   &Dt &DT &1     &DMtop &wDt   &3 \\
$b_L^1$   &Db &DB &1     &DMb   &wDb   &3 \\
$u_R^1$   &Su &SU &1     &SMu   &wSu   &3 \\
$d_R^1$   &Sd &SD &1     &SMd   &wSd   &3 \\
$c_R^1$   &Sc &SC &1     &SMc   &wSc   &3 \\
$s_R^1$   &Ss &SS &1     &SMs   &wSs   &3 \\
$t_R^1$   &St &ST &1     &SMtop &wSt   &3 \\
$b_R^1$   &Sb &SB &1     &SMb   &wSb   &3 \\
\hline
$u_L^2$   &$\sim Du$&$\sim DU$&1     &DMu2  &wDu2  &3\\
$d_L^2$   &$\sim Dd$&$\sim DD$&1     &DMd2  &wDd2  &3\\
$c_L^2$   &$\sim Dc$&$\sim DC$&1     &DMc2  &wDc2  &3\\
$s_L^2$   &$\sim Ds$&$\sim DS$&1     &DMs2  &wDs2  &3\\
$t_L^2$   &$\sim Dt$&$\sim DT$&1     &DMtop2&wDt2  &3\\
$b_L^2$   &$\sim Db$&$\sim DB$&1     &DMb2  &wDb2  &3\\
$u_R^2$   &$\sim Su$&$\sim SU$&1     &SMu2  &wSu2  &3\\
$d_R^2$   &$\sim Sd$&$\sim SD$&1     &SMd2  &wSd2  &3\\
$c_R^2$   &$\sim Sc$&$\sim SC$&1     &SMc2  &wSc2  &3\\
$s_R^2$   &$\sim Ss$&$\sim SS$&1     &SMs2  &wSs2  &3\\
$t_R^2$   &$\sim St$&$\sim ST$&1     &SMtop2&wSt2  &3\\
$b_R^2$   &$\sim Sb$&$\sim SB$&1     &SMb2  &wSb2  &3\\
\hline
\end{tabular}
\end{indented}
\end{table}

\subsection{Variables\label{sec:parameters}}

The input parameters for any given physics scenario are
defined in the {\tt varsN.mdl} model file.
In principle, MUED has only two additional input parameters beyond the SM:
the radius $R$ of the extra dimension and the cut-off scale $\Lambda$.
For convenience, we use the inverse radius $R^{-1}$ and the 
number of KK levels $\Lambda R$ which can fit below the scale $\Lambda$. 
$R^{-1}$ has dimensions of GeV, while $\Lambda R$ is dimensionless.
Our additions to the {\tt varsN.mdl} model file are listed in
Table~\ref{tab:parameter}. As seen from the table,
we also include several other variables of interest.
{\tt RG} is used to turn on and off the running of coupling constants,
while {\tt scaleN} is the renormalization scale $\mu$ at which the couplings 
are evaluated. The remaining parameters in Table~\ref{tab:parameter}
are some useful numerical constants related to the RGE running of the
gauge couplings (see Section~\ref{sec:running}).

\begin{table}[htb]
\caption{Parameters added to the {\tt varsN.mdl} model file. 
\label{tab:parameter}}
\begin{indented}
\item[] 
\centerline{
\begin{tabular}{cccl}
\hline \hline
Parameters	& Default values	& Symbols 	& Comments \\
\hline \hline
Rinv		& 500			& $R^{-1}$ 	& Inverse radius of the extra dimension \\
\hline
LR		& 20			& $\Lambda R$ 	& The number of KK levels below $\Lambda$\\
\hline
RG		& 1			&		& 1 turn on the running of the coupling constants  \\
		&			&		& 0 turn off the running of the coupling constants \\
\hline
		& 			& 		& Renormalization scale, $\mu = \frac{n}{R}$ \\
scaleN		& 2			& n		& n=2 can be used for KK level 1 pair production \\ 
                &                       &               & ~~~~~~~or level 2 single production \\
		&			&		& n=4 can be used for KK level 2 pair production \\
\hline
cb1		& 6.8333		& $b_1$		& $\frac{41}{6}$, The coefficient of the SM $\beta$-function for $U(1)_Y$ \\
cb2		& -3.16667		& $b_2$		& $-\frac{19}{6}$, The coefficient of the SM $\beta$-function for $SU(2)_W$ \\
cb3		& -7			& $b_3$		& $-7$, The coefficient of the SM $\beta$-function for $SU(3)_c$ \\
\hline
cb1t		& 6.8333		& $\tilde{b}_1$	& $\frac{41}{6}$, The coefficient of the KK $\beta$-function for $U(1)_Y$ \\
cb2t		& -2.83333		& $\tilde{b}_2$	& $-\frac{17}{6}$, The coefficient of the KK $\beta$-function for $SU(2)_W$ \\
cb3t		& -6.5			& $\tilde{b}_3$	& $-\frac{13}{2}$, The coefficient of the KK $\beta$-function for $SU(3)_c$ \\
c1MZ            & 98.4151               &$\alpha_1^{-1}$& $\alpha_1^{-1} (\mu=M_Z)$ \\
c2MZ            & 29.5846               &$\alpha_2^{-1}$& $\alpha_2^{-1} (\mu=M_Z)$ \\
c3MZ            & 8.53244               &$\alpha_3^{-1}$& $\alpha_3^{-1} (\mu=M_Z)$ \\
\hline
\end{tabular}
}
\end{indented}
\end{table}

\subsection{Constraints\label{sec:constraints}}

The {\tt funcN.mdl} model file is reserved for variables which are
not numerical inputs, but are instead computed in terms of the 
parameters already defined in the {\tt varsN.mdl} model file.
In our case, we use {\tt funcN.mdl} to
supply the masses and two-body decay widths of the KK particles
introduced in Section~\ref{sec:particles}. 
Therefore they are automatically computed by {\tt CalcHEP}/{\tt CompHEP} 
at the beginning of each numerical session.
The masses for all KK particles are
evaluated based on the 1-loop formulas of Ref.~\cite{Cheng:2002iz}
and we have also made numerical cross-checks with the results from the 
private code used in Ref.~\cite{Cheng:2002iz}. 
Our formulas for the widths have been derived analytically and cross-checked 
with {\tt CalcHEP}/{\tt CompHEP} (see Section~\ref{sec:pheno}).
A partial list of 2 body decay widths can be found 
in~\cite{Cheng:2002ab,Carone:2003ms,Datta:2005zs} and 
our formulas agree with their expressions. 
In the older versions of {\tt CalcHEP}/{\tt CompHEP},
defining the widths as constraints was very convenient in our implementation,
since one did not have to launch a separate numerical session 
for their calculation, and then enter their numerical values 
as input parameters. However, the more recent versions of 
{\tt CalcHEP} and {\tt CompHEP} allow for the automatic calculation of the 
particle widths on the fly, using
the interactions defined in the {\tt lgrngN.mdl} model file.
Our implementation thus allows for backward compatibility with older versions of 
{\tt CalcHEP}/{\tt CompHEP}. 

\subsection{Interactions}

The new interactions of the KK particles of Section~\ref{sec:particles}
are added to the {\tt lgrngN.mdl} model file.
We include the usual bulk interactions, as well as the
KK number violating boundary interactions 
listed in Table~\ref{tab:200}~\cite{Cheng:2002iz}.
Since the Weinberg angle $\theta_n$ for any $n\ge1$ 
is small~\cite{Cheng:2002iz}, we ignore the mixing 
among the neutral KK gauge bosons. Thus 
the KK-photon $\gamma_n$ is identical to the 
hypercharge gauge boson $B_n$ and the KK $Z$-boson $Z_n$ is 
identical to the neutral $SU(2)_W$ gauge boson $W_n^{3}$. 
We also ignore the mixing between $SU(2)_W$-doublet 
and $SU(2)_W$-singlet KK fermions. 

Our {\tt lgrngN.mdl} model file includes all interactions of level-1 
and level-2 KK particles except for the KK Higgs bosons. 
The phenomenology of the KK Higgs bosons is very model dependent,
depending on the value of the SM Higgs mass $m_h$ and the bulk 
Higgs mass term (see~\cite{Cheng:2002iz} for details). 
Therefore we omit any interactions involving KK Higgs bosons
\footnote{The collider phenomenology of the KK Higgs bosons has been discussed in \cite{Cembranos:2006gt,Bhattacherjee:2007wy,Bandyopadhyay:2009gd}.}.

The UED Lagrangian can be easily derived as shown in \ref{app:vertices}.
Here we only point out how to deal with 4-point interactions involving KK gluons,
since this case requires special treatment when 
implemented in {\tt CalcHEP}/{\tt CompHEP}.

The Lagrangian for the quartic interactions with KK gluons is the following
\be
\mathcal{L}_{4} &=& - \frac{1}{4} g_3^2 f^{abc} f^{ade} G^{0,b}_\mu G^{0,c}_\nu G^{0,d\mu} G^{0,e\nu} 
	- \frac{g_3^2}{2} f^{abc} f^{ade} G^{0,d}_\mu G^{0,e}_\nu G^{1,b\mu} G^{1,c\nu} \nonumber \\
	&-& \frac{g_3^2}{4} \left ( f^{abc} ( G^{0,b}_\mu G^{1,c}_\nu + G^{0,c}_\nu G^{1,b}_\mu ) \right )^2
	- \frac{1}{4}\cdot \frac{3}{2}g_3^2 f^{abc} f^{ade} G^{1,b}_\mu G^{1,c}_\nu G^{1,d\mu} G^{1,e\nu} \, .
\label{4g}
\ee
The color structure of these 4-point interactions cannot be directly 
written down in {\tt CalcHEP}/{\tt CompHEP} format.
Hence, to implement this vertex in {\tt CalcHEP}/{\tt CompHEP}, we use 
the following trick. We introduce three auxiliary tensor fields 
$t^a_{\mu\nu}$, $s^a_{\mu\nu}$ and $u^a_{\mu\nu}$
in the same way as the original {\tt CalcHEP}/{\tt CompHEP} approach for 
SM gluons. Then one can rewrite the Lagrangian as
\be
{\cal L} &=& -\frac{1}{2} t^a_{\mu\nu} t^{a\mu\nu} + i \frac{g_3}{\sqrt{2}} t^a_{\mu\nu} f^{abc} G^{0b\mu} G^{0c\nu}
		+ i \frac{g_3}{\sqrt{2}} t^a_{\mu\nu} f^{abc} G^{1b\mu} G^{1c\nu} \nonumber \\
	&& - \frac{1}{2} s^a_{\mu\nu} s^{a\mu\nu} + i \frac{g_3}{2} s^a_{\mu\nu} f^{abc} G^{1b\mu} G^{1c\nu} \nonumber \\
	&& - \frac{1}{2} u^a_{\mu\nu} u^{a\mu\nu} +
		i \frac{g_3}{\sqrt{2}} u^a_{\mu\nu} f^{abc} \left ( G^{0b\mu} G^{1c\nu} + G^{1b\mu} G^{0c\nu} \right ) \nonumber\\
	&=& -\frac{1}{2} \left ( t^a_{\mu\nu} - i g_3 \frac{1}{\sqrt{2}} f^{abc} G^{0b}_\mu G^{0c}_\nu -i g_3 \frac{1}{\sqrt{2}} f^{abc} G^{1b}_\mu G^{1c}_\nu \right )^2 \\
		&& -\frac{1}{4} g_3^2  f^{abc}  f^{ade} \left ( G^{0b}_\mu G^{0c}_\nu + G^{1b}_\mu G^{1c}_\nu \right ) \left ( G^{0d\mu} G^{0e\nu} + G^{1d\mu} G^{1e\nu} \right ) \nonumber\\
	&&- \frac{1}{2} \left ( s^a_{\mu\nu} - i g_3 \frac{1}{{2}}f^{abc} G^{1b}_\mu G^{1c}_\nu \right )^2
		- \frac{1}{8} g_3^2 f^{abc} f^{ade} G^{1b}_\mu G^{1c}_\nu G^{1d\mu} G^{1e\nu}  \nonumber\\
	&&- \frac{1}{2} \left ( u^a_{\mu\nu} - i g_3 \frac{1}{\sqrt{2}}f^{abc} \left ( G^{0b}_\mu G^{1c}_\nu  + G^{1b}_\mu G^{0c}_\nu\right ) \right )^2 \nonumber \\
	&& -\frac{1}{4} g_3^2 \left ( f^{abc} \left ( G^{0b}_\mu G^{1c}_\nu  + G^{1b}_\mu G^{0c}_\nu \right ) \right )^2 \, ,\nonumber
\ee
It is easy to show that
the functional integration over the three auxiliary tensor fields reproduces 
the 4-gluon interactions (\ref{4g}). 

\subsection{Running of the coupling constants}
\label{sec:running}
Due to the additional contributions from the KK modes to the beta functions, 
the gauge couplings run faster in theories with extra dimensions.
The RGE for $\alpha_i\equiv \frac{g_i^2}{4\pi}$ is given by~\cite{Dienes:1998vg}
\be
\frac{d \alpha^{-1}_i}{d t} = - \frac{b_i - \tilde{b}_i}{2\pi} 
     - \frac{\tilde{b}_i X_{\delta}}{2\pi} \left ( \frac{\mu}{\mu_0} \right )^{\delta} \, ,
\label{gRGE}
\ee
where $\delta$ is the number of extra dimensions, $\mu_0$ is some reference energy scale,
$X_{\delta} = \frac{2\pi^{\delta/2}}{\delta \Gamma (\delta/2)}$,
\be
\left( b_1 , b_2 , b_3 \right) = \left(\frac{41}{6},-\frac{19}{6},-7\right)
\ee
are the SM beta function coefficients, while 
\be
\left( \tilde{b}_1 , \tilde{b}_2 , \tilde{b}_3 \right) = \left(\frac{41}{6},-\frac{17}{6},-\frac{13}{2}\right) 
\ee
correspond to the contributions of the Kaluza-Klein states 
at each massive KK excitation level~\cite{Perez-Lorenzana:1999qb,Cheng:1999fu}.
The solution to (\ref{gRGE}) becomes
\be
\alpha_i^{-1} = \alpha^{-1}_i (M_Z ) - \frac{b_i}{2\pi} \ln\frac{\mu}{M_Z} + \frac{\tilde{b}_i}{2\pi} \ln\frac{\mu}{\mu_0}
	- \frac{\tilde{b}_i X_{\delta}}{2\pi \delta} \left [ \left ( \frac{\mu}{\mu_0} \right )^{\delta} -1 \right ] \, .
\label{gRGEsol}
\ee
The effect of the RGE running (\ref{gRGEsol}) can be accounted for by setting the
{\tt RG} parameter in Table~\ref{tab:parameter} to 1 and
choosing the appropriate renormalization scale via 
{\tt scaleN}.

\section{Discussion}
\label{sec:pheno}

\subsection{Code validation}

In general, the availability of {\tt CalcHEP}/{\tt CompHEP} 
model files opens the door to a number of applications related to
collider phenomenology and dark matter searches.
Each such individual study contributes to the validation of
the code. Further consistency checks are provided by comparing to
existing analytical and/or numerical results in the literature.

\begin{figure}[tb]
\centerline{
\includegraphics[width=.47\linewidth]{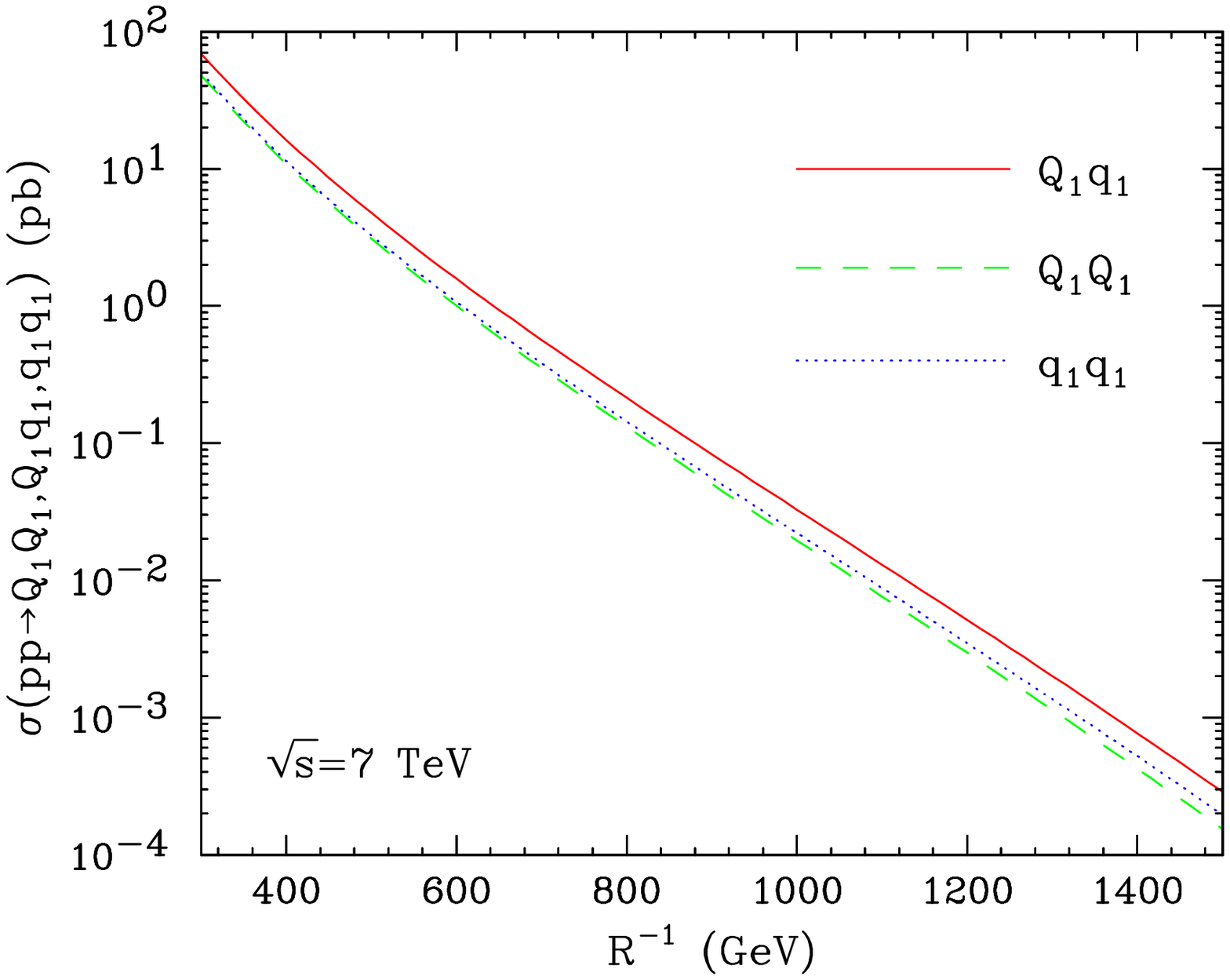} \hspace{0.1cm}
\includegraphics[width=.47\linewidth]{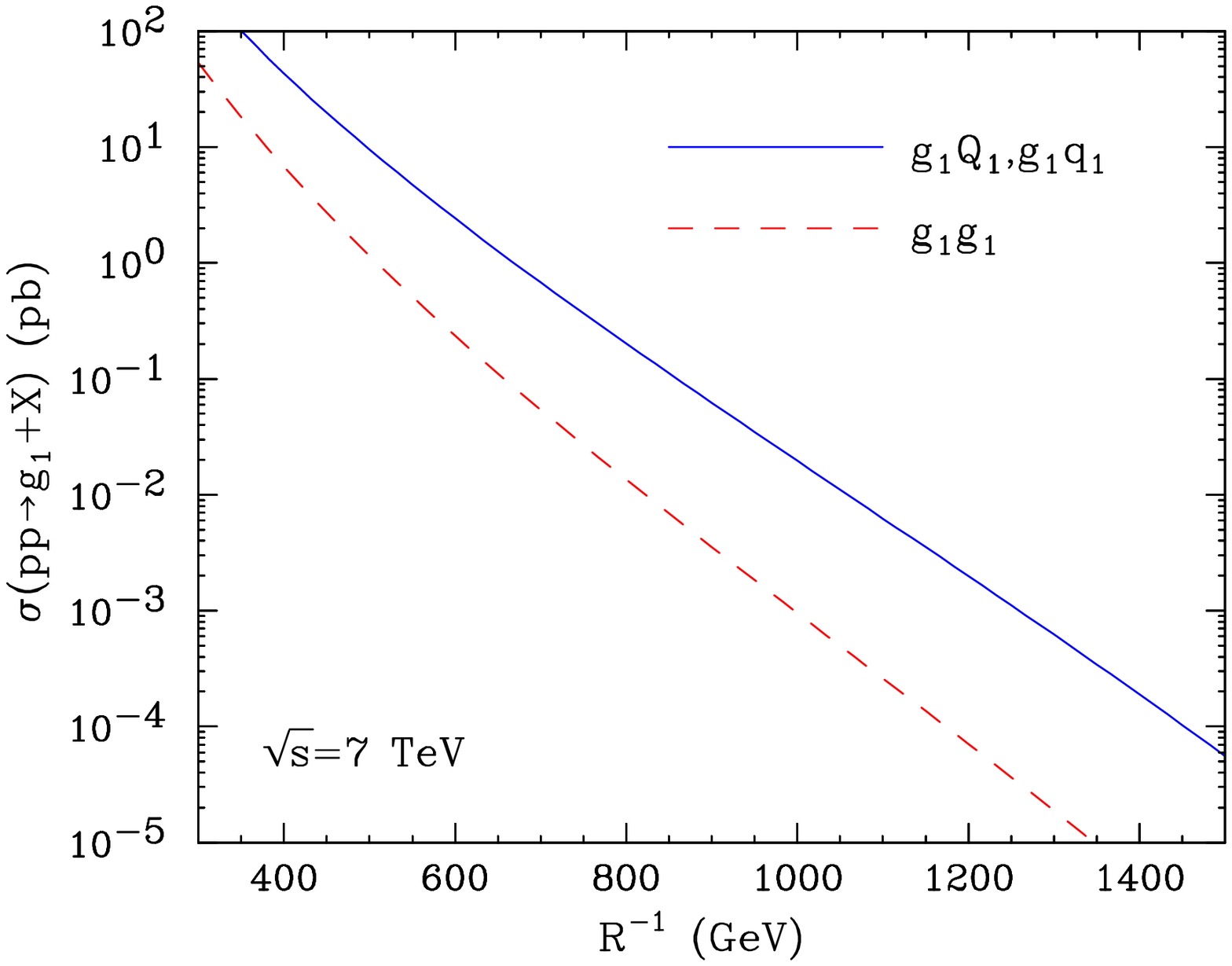}
}
\caption{Strong production of $n=1$ KK particles at the LHC for $\sqrt{s}=7$ TeV: 
(a) KK-quark pair production; (b) KK-quark/KK-gluon associated production and KK-gluon pair production.
The cross sections have been summed over all quark flavors and also include charge-conjugated contributions 
such as $Q_1 \bar{q}_1$, $\bar{Q}_1 q_1$, $g_1 \bar{Q}_1$, etc. 
We use CTEQ6L parton distributions 
\cite{Nadolsky:2008zw} and choose the scale of the strong coupling constant
$\alpha_s$ to be equal to the parton level center of mass energy.}
\label{fig:sigma_q1g1}
\end{figure}

\begin{itemize}
\item For starters, we have compared the KK mass spectrum 
calculated with our implementation 
to the results shown in Fig.~\ref{fig:spectrum}, which were obtained
independently in Ref.~\cite{Cheng:2002iz}. Using identical inputs, and 
neglecting the running of the gauge couplings (as was done in \cite{Cheng:2002iz}),
we found perfect agreement.
\item The interaction vertices of \ref{app:vertices} can be independently
derived with the automated tool {\tt LanHEP}~\cite{Semenov:2002jw}.
We checked some of the more technically challenging cases
(especially the self-interactions of gauge bosons) 
and also found agreement.
\item To minimize the possibility of typing mistakes, we
computed analytically the cross-sections for a selected 
number of simple scattering processes,
and compared to the analytical expressions derived by 
{\tt CalcHEP}/{\tt CompHEP}.
\item We have similarly checked that the KK particle widths 
calculated from our analytical expressions agree with those
computed with {\tt CalcHEP}/{\tt CompHEP} by means of our MUED
implementation.
\item Our analytic formulas for decay widths agree with the 
expressions given in~\cite{Cheng:2002ab,Carone:2003ms,Datta:2005zs}.
\item Our implementation was used for the analytic calculation of 
{\em all} (co)annihilation cross-sections of level 1 KK particles~\cite{Kong:2005hn}
and the results were in complete agreement 
with~\cite{Servant:2002aq,Burnell:2005hm}. 
\item Our model files have already been used for various collider 
studies~\cite{Datta:2005vx,Kong:2006pi,Buescher:2006jm,Datta:2005zs,Datta:2005gm,
Battaglia:2005zf,Battaglia:2005ma,Kong:2005hn,Accomando:2004sz,Konar:2009ae}. One example is shown 
in Fig.~\ref{fig:sigma_q1g1}, which shows the strong production 
cross-section of level 1 KK particles at the imminent LHC energy of 7 TeV.
\item We have compared results for various production cross-sections
in MUED to those in published papers~\cite{Rizzo:2001sd,Macesanu:2002db}
and find agreement.
\item Our model files were also cross-checked against 
the known analytical expressions for various invariant mass distributions
\cite{Smillie:2005ar,Burns:2008cp,Kong:2010mh}.
\item Our model files have also been tested by other groups, for example in
creating Pythia\underline{ }UED~\cite{Skands:2005vi,Allanach:2006fy,ElKacimi:2009zj}, which
implemented the matrix elements for certain processes in {\tt PYTHIA}~\cite{Sjostrand:2003wg}. 
Another extensive comparison to an independent MUED implementation 
via FeynRules was done in \cite{Christensen:2009jx}.
\end{itemize}

\subsection{Outlook}

Moving forward, it is important to be mindful of the 
limitations of our implementation. First of all, it is still
Minimal UED, and the spectrum is quite constrained, given in 
terms of only 2 parameters: $R^{-1}$ and $\Lambda$.
If a signal consistent with UED is discovered at the LHC or the Tevatron, 
one would like to start testing the data with a more general UED framework, which
allows for the presence of arbitrary boundary terms at the scale $\Lambda$.
Work along these lines has already started and a beta version
of the corresponding UED model files is available from the authors upon request.

\section*{Acknowledgments}
We are grateful to Priscila de Aquino, Neil Christensen and Claude Duhr
for independent extensive testing of our model files against the results from
{\tt FeynRules}, in the process of which a typo in the original version of our 
MUED model files was uncovered.
AD is partially supported by funding available from the Department of Atomic Energy,
Government of India, for the Regional Centre for Accelerator-based Particle Physics,
Harish-Chandra Research Institute.
KK is supported in part by DOE under contract DE-AC02-76SF00515.
KM is supported in part by a US Department of Energy grant DE-FG02-97ER41029.

\newpage
\appendix
\section{UED Lagrangian in 5 dimensions}
\label{app:5d}
\renewcommand{\theequation}{A.\arabic{equation}}
\renewcommand{\thetable}{A.\arabic{table}}
\setcounter{equation}{0}
\setcounter{table}{0}


The Lagrangian for the 5-dimensional UED model is written as
\be
{\cal L}_{Gauge} &=& \intez \left \{-\frac{1}{4} B_{MN} B^{MN} -\frac{1}{4} W^a_{MN} W^{aMN} -\frac{1}{4} G^A_{MN} G^{AMN} \right \}\, , \label{Lgauge}\\
{\cal L}_{GF}    &=& \intez \left \{ -\frac{1}{2\xi} \left ( \partial^\mu B_\mu - \xi \partial_5 B_5 \right )^2  
                                 -\frac{1}{2\xi} \left ( \partial^\mu W^a_\mu - \xi \partial_5 W^a_5 \right )^2 \right. \label{LGF}\\
&& \hspace*{2cm} \left. -\frac{1}{2\xi} \left ( \partial^\mu G^A_\mu - \xi \partial_5 G^G_5 \right )^2 \right \} \nonumber\, ,\\
{\cal L}_{Leptons} &=& \intez \left \{ i \bar{L}(x,y) \Gamma^M D_M L (x,y) + i \bar{e}(x,y) \Gamma^M D_M e(x,y) \right \} \, , \\
{\cal L}_{Quarks} &=& \intez \left \{ i \bar{Q}(x,y) \Gamma^M D_M Q(x,y) + i \bar{u}(x,y) \Gamma^M D_M u(x,y) \right. \\
                 && \hspace*{6.cm} \left.+ i \bar{d}(x,y) \Gamma^M D_M d(x,y) \right \} \nonumber\, ,\\
{\cal L}_{Yukawa} &=& \intez \left \{ \lambda_u \bar{Q}(x,y) u(x,y)i\tau^2 H^*(x,y) + \lambda_d \bar{Q}(x,y) d(x,y) H(x,y) \right. \nonumber\\
        && \left. + \lambda_e \bar{L}(x,y) e(x,y) H(x,y) \right \}  \, ,\\
{\cal L}_{Higgs} &=& \intez \left [ \left ( D_M H(x,y) \right )^\dagger \left ( D^M H(x,y) \right ) + \mu^2 H^\dagger (x,y)
	H(x,y)\right.  \nonumber\\
     && \hspace{3cm}\left. - \lambda \left ( H^\dagger (x,y) H(x,y) \right )^2 \right ] \, , \label{LHiggs}
\ee
in terms of 5-dimensional fields decomposed as discussed in Section~\ref{sec:dec}:

\be
H(x, y) &=& \frac{1}{\sqrt{\pi R}} \left \{ H(x) + \sqrt{2} \summ  H_n (x) \cos (\frac{ny}{R})  \right  \}  \nonumber \, ,\\
B_\mu (x,y) &=& \frac{1}{\sqrt{\pi R}} \left \{ B_\mu^{0}(x) + \sqrt{2} \summ B_\mu^{n} (x) \cos (\frac{ny}{R}) \right  \} \nonumber \, ,\\
B_5 (x,y) &=& \sqrt{\frac{2}{\pi R}} \summ B_5^n (x) \sin (\frac{n y}{R}) \nonumber\, ,\\
W_\mu (x,y) &=& \frac{1}{\sqrt{\pi R}} \left \{ W_\mu^{0}(x) + \sqrt{2} \summ W_\mu^{n} (x) \cos (\frac{ny}{R}) \right  \} \nonumber\, ,\\
W_5 (x,y) &=& \sqrt{\frac{2}{\pi R}} \summ W_5^n (x) \sin (\frac{n y}{R}) \nonumber\, ,\\
G_\mu (x,y) &=& \frac{1}{\sqrt{\pi R}} \left \{ G_\mu^{0}(x) + \sqrt{2} \summ G_\mu^{n} (x) \cos (\frac{ny}{R}) \right  \} \, , \label{expansions} \\
G_5 (x,y) &=& \sqrt{\frac{2}{\pi R}} \summ G_5^n (x) \sin (\frac{n y}{R}) \nonumber\, ,\\
Q(x, y) &=& \frac{1}{\sqrt{\pi R}} \left \{ q_L (x) + \sqrt{2} \summ  \left [ P_L Q_L^n (x) \cos (\frac{ny}{R}) + P_R Q_R^n (x) \sin (\frac{ny}{R} ) \right ] \right \} \nonumber\, ,\\
u(x, y) &=& \frac{1}{\sqrt{\pi R}} \left \{ u_R (x) + \sqrt{2} \summ  \left [ P_R u_R^n (x) \cos (\frac{ny}{R}) + P_L u_L^n (x) \sin (\frac{ny}{R} ) \right ] \right \} \nonumber\, ,\\
d(x, y) &=& \frac{1}{\sqrt{\pi R}} \left \{ d_R (x) + \sqrt{2} \summ  \left [ P_R d_R^n (x) \cos (\frac{ny}{R}) + P_L d_L^n (x) \sin (\frac{ny}{R} ) \right ] \right \} \nonumber\, ,\\
L(x, y) &=& \frac{1}{\sqrt{\pi R}} \left \{ L_L (x) + \sqrt{2} \summ  \left [ P_L L_L^n (x) \cos (\frac{ny}{R}) + P_R L_R^n (x) \sin (\frac{ny}{R} ) \right ] \right \} \nonumber\, ,\\
e(x, y) &=& \frac{1}{\sqrt{\pi R}} \left \{ e_R (x) + \sqrt{2} \summ  \left [ P_R e_R^n (x) \cos (\frac{ny}{R}) + P_L e_L^n (x) \sin (\frac{ny}{R} ) \right ] \right \} \nonumber \, .
\ee
Here $H(x,y)$ is the 5D Higgs scalar field and $(B_\mu (x,y), B_5 (x,y))$, 
$(W_\mu (x,y), W_5 (x,y))$ and $(G_\mu (x,y), G_5 (x,y))$ are the
5D gauge fields $B_M$, $W_M$ and $G_M$
for $U(1)_Y$, $SU(2)_W$ and $SU(3)_c$, respectively. 
The 5D index $M$ runs over $M=\mu,5$, where $\mu=0,1,2,3$.
The $SU(2)_W$ and $SU(3)_c$ gauge fields are
\ben
W_M &\equiv& W_M^a \frac{\tau^a}{2} \, ,\\
G_M &\equiv& G_M^A \frac{\lambda^A}{2} \, ,
\een
where $\tau^a$, $a=1,2,3$, are the usual Pauli matrices and 
$\lambda^A$, $A=1,2,...,8$, are the usual Gell-Mann matrices.
The 5D field strength tensors for $U(1)_Y$, $SU(2)_W$ and $SU(3)_c$ are defined as follows
\be
B_{MN} &=& \partial_M B_N - \partial_N B_M \nonumber \, , \\
W^a_{MN} &=& \partial_M W^a_N - \partial_N W^a_M + g_2^{(5)} \epsilon^{abc}W^b_M W^c_N \, , \\
G^A_{MN} &=& \partial_M G^A_N - \partial_N G^A_M + g_3^{(5)} f^{ABC} G^B_M G^C_N  \nonumber \, ,
\ee
where $\epsilon^{abc}$ and $f^{ABC}$ are the structure constants for $SU(2)_W$ and $SU(3)_c$, respectively.
The parameter $\xi$ in (\ref{LGF}) is the gauge fixing parameter in the generalized $R_\xi$ gauge.

The 5-dimensional (4-dimensional) gauge couplings are denoted by $g_i^{(5)}$ ($g_i$), where
$i=1,2,3$ stands for $U(1)_Y$, $SU(2)_W$ and $SU(3)_c$, correspondingly. 
The two types of couplings are related by 
\be
g_i = \frac{g^{(5)}_i}{\sqrt{\pi R}} \, .
\ee

Finally, $Q(x,y)$ and $L(x,y)$ are the $SU(2)_W$-doublet fermions from Table~\ref{tab:fermion},
while $u(x,y)$, $d(x,y)$ and $e(x,y)$ are the corresponding $SU(2)_W$-singlet fermions
from Table~\ref{tab:fermion}. $P_{L,R} = \frac{1 \mp \g^5}{2}$ are the 4D chiral projectors
in terms of the usual $\gamma_5$ matrix. The gamma matrices in 5D
\be
\Gamma^M &=& (\g^\mu,  i\g^5)\, ,
\ee
satisfy the Dirac-Clifford algebra
\be
\{ \Gamma^M , \Gamma^N \} = 2 g^{MN}\, ,
\ee
where $g^{MN}$ is the 5D metric
\be
g_{MN} &=& \left(
\begin{array}{cc}
g_{\mu\nu} & 0 \\
0          & -1
\end{array} \right) \, ,
\ee
and $g^{\mu\nu} =(+---)$ is the usual 4D metric.

The covariant derivatives act on 5D fields as follows
\be
D_M Q (x,y) &=& \left ( \partial_M + i {g_3^{(5)}} G_M + i {g_2^{(5)}} W_M + i \frac{Y_Q}{2} {g_1^{(5)}} B_M \right ) Q(x,y) \, ,\\
D_M u (x,y) &=& \left ( \partial_M + i {g_3^{(5)}} G_M                     + i \frac{Y_u}{2} {g_1^{(5)}} B_M \right ) u(x,y) \nonumber\, ,\\
D_M d (x,y) &=& \left ( \partial_M + i {g_3^{(5)}} G_M                     + i \frac{Y_d}{2} {g_1^{(5)}} B_M \right ) d(x,y)\nonumber \, ,\\
D_M L (x,y) &=& \left ( \partial_M                     + i {g_2^{(5)}} W_M + i \frac{Y_L}{2} {g_1^{(5)}} B_M \right ) L(x,y) \nonumber\, ,\\
D_M e (x,y) &=& \left ( \partial_M                                         + i \frac{Y_e}{2} {g_1^{(5)}} B_M \right ) e(x,y) \nonumber\, ,
\ee
where the fermion hypercharges are 
$Y_Q=\frac{1}{3}$,
$Y_u=\frac{4}{3}$,
$Y_d=-\frac{2}{3}$,
$Y_L=-1$ and
$Y_e=-2$.

It is now a rather straightforward but tedious exercise to substitute
the expansions (\ref{expansions}) into the 5D Lagrangians (\ref{Lgauge}-\ref{LHiggs})
and perform the integration over $y$ with the help of the orthonormality relations 
listed in \ref{app:ortho}. The resulting Feynman rules in terms of 4-dimensional 
fields are listed in \ref{app:vertices}.

\newpage
\section{Orthonormality Relations}
\label{app:ortho}

\renewcommand{\theequation}{B.\arabic{equation}}
\renewcommand{\thetable}{B.\arabic{table}}
\setcounter{equation}{0}
\setcounter{table}{0}

The following orthonormality relations can be used in the process of
compactifying the 5-dimensional Lagrangian listed in \ref{app:5d}.
\be
\int_{0}^{\pi R}d y \cos (\frac{m y}{R}) \cos (\frac{n y}{R}) &=& \frac{\pi R}{2} \delta_{m,n} \nonumber \, ,\\
\int_{0}^{\pi R}d y \sin (\frac{m y}{R}) \sin (\frac{n y}{R}) &=& \frac{\pi R}{2} \delta_{m,n} \nonumber \, ,\\
\int_{0}^{\pi R}d y \cos (\frac{m y}{R}) \cos (\frac{n y}{R}) \cos (\frac{l y}{R})  &=& \frac{\pi R}{4} \Delta_{mnl}^1 \nonumber \, ,\\
\int_{0}^{\pi R}d y \cos (\frac{m y}{R}) \cos (\frac{n y}{R}) \cos (\frac{l y}{R}) \cos (\frac{k y}{R})  &=& \frac{\pi R}{8} \Delta_{mnlk}^2 \nonumber \, ,\\
\int_{0}^{\pi R}d y \sin (\frac{m y}{R}) \sin (\frac{n y}{R}) \sin (\frac{l y}{R}) \sin (\frac{k y}{R})  &=& \frac{\pi R}{8} \Delta_{mnlk}^3 \nonumber  \, ,\\
\int_{0}^{\pi R}d y \sin (\frac{m y}{R}) \sin (\frac{n y}{R}) \cos (\frac{l y}{R})  &=& \frac{\pi R}{4} \Delta_{mnl}^4  \, ,\\
\int_{0}^{\pi R}d y \sin (\frac{m y}{R}) \sin (\frac{n y}{R}) \cos (\frac{l y}{R}) \cos (\frac{k y}{R})  &=& \frac{\pi R}{8} \Delta_{mnlk}^5 \nonumber \, ,\\
\int_{0}^{\pi R}d y \cos (\frac{m y}{R}) \sin (\frac{n y}{R}) &=& 0 \nonumber \, ,\\
\int_{0}^{\pi R}d y \sin (\frac{m y}{R}) \sin (\frac{n y}{R}) \sin (\frac{l y}{R})  &=& 0  \nonumber \, ,\\
\int_{0}^{\pi R}d y \sin (\frac{m y}{R}) \cos (\frac{n y}{R}) \cos (\frac{l y}{R})  &=& 0  \nonumber \, ,\\
\int_{0}^{\pi R}d y \sin (\frac{m y}{R}) \cos (\frac{n y}{R}) \cos (\frac{l y}{R})  \cos (\frac{k y}{R}) &=& 0 \nonumber  \, ,\\
\int_{0}^{\pi R}d y \sin (\frac{m y}{R}) \sin (\frac{n y}{R}) \sin (\frac{l y}{R})  \cos (\frac{k y}{R}) &=& 0\nonumber  \, ,
\ee
where the $\Delta$ symbols are defined as
\be
\Delta_{mnl}^1  &=& \delta_{l,m+n} + \delta_{n,l+m} + \delta_{m,l+n}  \, ,\\
\Delta_{mnlk}^2 &=& \delta_{k,l+m+n} + \delta_{l,m+n+k} + \delta_{m,n+k+l}+ \delta_{n,k+l+m} \\
               && +\delta_{k+m,l+n} + \delta_{k+l,m+n} +\delta_{k+n,l+m} \nonumber\, ,\\
\Delta_{mnlk}^3 &=& - \delta_{k,l+m+n} - \delta_{l,m+n+k} - \delta_{m,n+k+l} -\delta_{n,k+l+m} \\
               &&+\delta_{k+l,m+n} + \delta_{k+m,l+n} +\delta_{k+n,l+m} \nonumber \, ,\\
\Delta_{mnl}^4  &=& -\delta_{l,m+n} + \delta_{n,l+m} + \delta_{m,l+n} \, ,\\
\Delta_{mnlk}^5 &=& - \delta_{k,l+m+n} - \delta_{l,m+n+k} + \delta_{m,n+k+l}+ \delta_{n,k+l+m} \\
               && - \delta_{k+l,m+n} + \delta_{k+m,l+n} +\delta_{k+n,l+m} \nonumber \, .
\ee

\newpage
\section{Feynman Rules}
\label{app:vertices}

Here we list some of the KK-number conserving vertices with subscripts `n' standing for the KK-level, 
which are obtained after compactifying the 5-dimensional Lagrangian of \ref{app:5d}
with the help of the orthonormality relations of \ref{app:ortho}.
For KK-number violating (but still KK-parity conserving) vertices, refer to 
Fig. \ref{fig:200} and Table \ref{tab:200}. \\

\begin{center}
\begin{tabular}{lr}
\hspace*{-3cm} \input{G-q1-q1.pstex_t} 		& \hspace*{3cm} \input{G2-q1-q1.pstex_t} \\ \\
\hspace*{-3cm} \input{G1-qd1-q0.pstex_t} 	& \hspace*{3cm} \input{G1-qs1-q0.pstex_t}\\ \\
\hspace*{-3cm} \input{G-G1-G1.pstex_t}		& \hspace*{3cm} \\ \\
\hspace*{-3cm} \input{G-G-G1-G1.pstex_t}	& \hspace*{3cm} \\ \\
\end{tabular}
\end{center}

\newpage
\begin{center}
\begin{tabular}{lr}
& \\ \\
\hspace*{-3cm} \input{G1-G1-G1-G1.pstex_t}      & \\ \\
\hspace*{-3cm} \input{A-f1-f1.pstex_t} 		& \hspace*{4cm} \input{Z-fd1-fd1.pstex_t} \\ \\
\hspace*{-3cm} \input{Z-fs1-fs1.pstex_t}	& \hspace*{4cm} \input{W-fd1-fd1.pstex_t} \\ \\
\hspace*{-3cm} \input{B1-fd1-fd1.pstex_t} 	& \hspace*{4cm} \input{B1-fs1-fs1.pstex_t} \\ \\
\hspace*{-3cm} \input{Z1-fd1-f0.pstex_t} 	& \hspace*{4cm} \input{W1-fd1-f0.pstex_t} \\ \\
\end{tabular}
\end{center}

\newpage

\begin{center}
\begin{tabular}{lr}
& \\ 
\hspace*{-3cm} \input{B2-fd1-fd1.pstex_t} 	& \hspace*{4cm} \input{B2-fs1-fs1.pstex_t} \\ \\
\hspace*{-3cm} \input{Z2-fd1-fd1.pstex_t} 	& \hspace*{4cm} \input{W2-fd1-fd1.pstex_t} \\ \\
\end{tabular}
\end{center}
\begin{center}
\begin{tabular}{c}
\hspace*{-8cm}\input{A-W1-W1.pstex_t} \\ \\
\hspace*{-8cm}\input{Z1-W-W1.pstex_t} \\ \\
\hspace*{-8cm}\input{Z-W1-W1.pstex_t} \\ \\
\end{tabular}
\end{center}

\newpage
\begin{center}
\begin{tabular}{c}
\\ 
\hspace*{-8cm}\input{Z2-W1-W1.pstex_t} \\ \\
\hspace*{-8cm}\input{A-A-W1-W1.pstex_t} \\ \\
\hspace*{-8cm}\input{A-W-W1-Z1.pstex_t} \\ \\
\end{tabular}
\end{center}
\begin{center}
\begin{tabular}{c}
\hspace*{-8cm}\input{A-W1-W1-Z.pstex_t} \\ \\
\hspace*{-8cm}\input{A-W1-W1-Z2.pstex_t} \\ \\
\end{tabular}
\end{center}
\newpage

\begin{center}
\begin{tabular}{c}
\\ 
\hspace*{-8cm}\input{W-W-W1-W1.pstex_t} \\ \\
\hspace*{-8cm}\input{W-W-Z1-Z1.pstex_t} \\ \\
\hspace*{-8cm}\input{W-W1-Z-Z1.pstex_t} \\ \\
\hspace*{-8cm}\input{W-W1-Z1-Z2.pstex_t} \\ \\
\end{tabular}
\end{center}
\begin{center}
\begin{tabular}{c}
\hspace*{-8cm}\input{W1-W1-W1-W1.pstex_t} \\ \\
\end{tabular}
\end{center}
\newpage

\begin{center}
\begin{tabular}{c}
\hspace*{-8cm}\input{W1-W1-Z-Z.pstex_t} \\ \\
\hspace*{-8cm}\input{W1-W1-Z-Z2.pstex_t} \\ \\
\hspace*{-8cm}\input{W1-W1-Z1-Z1.pstex_t} \\ \\
\hspace*{-8cm}\input{W1-W1-Z2-Z2.pstex_t} \\ \\
\end{tabular}
\end{center}


\end{document}

%% file: 200.pstex_t
\begin{picture}(0,0)%
\includegraphics{200.pstex}%
\end{picture}%
\setlength{\unitlength}{3947sp}%
\begingroup\makeatletter\ifx\SetFigFont\undefined%
\gdef\SetFigFont#1#2#3#4#5{%
  \reset@font\fontsize{#1}{#2pt}%
  \fontfamily{#3}\fontseries{#4}\fontshape{#5}%
  \selectfont}%
\fi\endgroup%
\begin{picture}(5505,1611)(286,-964)
\put(4351,-586){\makebox(0,0)[lb]{\smash{{\SetFigFont{12}{14.4}{\rmdefault}{\mddefault}{\updefault}{\color[rgb]{0,0,0}$f_1$}%
}}}}
\put(5101,-211){\makebox(0,0)[lb]{\smash{{\SetFigFont{12}{14.4}{\rmdefault}{\mddefault}{\updefault}{\color[rgb]{0,0,0}$V_1$}%
}}}}
\put(301,-136){\makebox(0,0)[lb]{\smash{{\SetFigFont{12}{14.4}{\rmdefault}{\mddefault}{\updefault}{\color[rgb]{0,0,0}$V_2$}%
}}}}
\put(2026,-811){\makebox(0,0)[lb]{\smash{{\SetFigFont{12}{14.4}{\rmdefault}{\mddefault}{\updefault}{\color[rgb]{0,0,0}$\bar{f}_0$}%
}}}}
\put(2101,464){\makebox(0,0)[lb]{\smash{{\SetFigFont{12}{14.4}{\rmdefault}{\mddefault}{\updefault}{\color[rgb]{0,0,0}$f_0$}%
}}}}
\put(3226,-211){\makebox(0,0)[lb]{\smash{{\SetFigFont{12}{14.4}{\rmdefault}{\mddefault}{\updefault}{\color[rgb]{0,0,0}$V_2$}%
}}}}
\put(4351, 89){\makebox(0,0)[lb]{\smash{{\SetFigFont{12}{14.4}{\rmdefault}{\mddefault}{\updefault}{\color[rgb]{0,0,0}$f_1$}%
}}}}
\put(5776,-886){\makebox(0,0)[lb]{\smash{{\SetFigFont{12}{14.4}{\rmdefault}{\mddefault}{\updefault}{\color[rgb]{0,0,0}$\bar{f}_0$}%
}}}}
\put(5776,464){\makebox(0,0)[lb]{\smash{{\SetFigFont{12}{14.4}{\rmdefault}{\mddefault}{\updefault}{\color[rgb]{0,0,0}$f_0$}%
}}}}
\end{picture}%

%% file: G-q1-q1.pstex_t
\begin{picture}(0,0)%
\includegraphics{G-q1-q1.pstex}%
\end{picture}%
\setlength{\unitlength}{3947sp}%
\begingroup\makeatletter\ifx\SetFigFont\undefined%
\gdef\SetFigFont#1#2#3#4#5{%
  \reset@font\fontsize{#1}{#2pt}%
  \fontfamily{#3}\fontseries{#4}\fontshape{#5}%
  \selectfont}%
\fi\endgroup%
\begin{picture}(1530,1669)(-14,-809)
\put(1351,689){\makebox(0,0)[lb]{\smash{{\SetFigFont{12}{14.4}{\rmdefault}{\mddefault}{\updefault}{\color[rgb]{0,0,0}$q^b_n$}%
}}}}
\put(1501,-136){\makebox(0,0)[lb]{\smash{{\SetFigFont{12}{14.4}{\rmdefault}{\mddefault}{\updefault}{\color[rgb]{0,0,0}$=-ig_3\g^\mu T^c_{ba}$}%
}}}}
\put(  1,-61){\makebox(0,0)[lb]{\smash{{\SetFigFont{12}{14.4}{\rmdefault}{\mddefault}{\updefault}{\color[rgb]{0,0,0}$G^c$}%
}}}}
\put(1351,-736){\makebox(0,0)[lb]{\smash{{\SetFigFont{12}{14.4}{\rmdefault}{\mddefault}{\updefault}{\color[rgb]{0,0,0}$\bar{q}^a_n$}%
}}}}
\end{picture}%

%% file: G2-q1-q1.pstex_t
\begin{picture}(0,0)%
\includegraphics{G2-q1-q1.pstex}%
\end{picture}%
\setlength{\unitlength}{3947sp}%
\begingroup\makeatletter\ifx\SetFigFont\undefined%
\gdef\SetFigFont#1#2#3#4#5{%
  \reset@font\fontsize{#1}{#2pt}%
  \fontfamily{#3}\fontseries{#4}\fontshape{#5}%
  \selectfont}%
\fi\endgroup%
\begin{picture}(1530,1669)(-14,-809)
\put(1501,-136){\makebox(0,0)[lb]{\smash{{\SetFigFont{12}{14.4}{\rmdefault}{\mddefault}{\updefault}{\color[rgb]{0,0,0}$=-i \frac{g_3}{\sqrt{2}} \g^\mu T^c_{ba}$}%
}}}}
\put(  1,-61){\makebox(0,0)[lb]{\smash{{\SetFigFont{12}{14.4}{\rmdefault}{\mddefault}{\updefault}{\color[rgb]{0,0,0}$G^c_2$}%
}}}}
\put(1351,689){\makebox(0,0)[lb]{\smash{{\SetFigFont{12}{14.4}{\rmdefault}{\mddefault}{\updefault}{\color[rgb]{0,0,0}$q^b_1$}%
}}}}
\put(1351,-736){\makebox(0,0)[lb]{\smash{{\SetFigFont{12}{14.4}{\rmdefault}{\mddefault}{\updefault}{\color[rgb]{0,0,0}$\bar{q}^a_1$}%
}}}}
\end{picture}%

%% file: G1-qd1-q0.pstex_t
\begin{picture}(0,0)%
\includegraphics{G1-qd1-q0.pstex}%
\end{picture}%
\setlength{\unitlength}{3947sp}%
\begingroup\makeatletter\ifx\SetFigFont\undefined%
\gdef\SetFigFont#1#2#3#4#5{%
  \reset@font\fontsize{#1}{#2pt}%
  \fontfamily{#3}\fontseries{#4}\fontshape{#5}%
  \selectfont}%
\fi\endgroup%
\begin{picture}(1530,1669)(-14,-809)
\put(1351,689){\makebox(0,0)[lb]{\smash{{\SetFigFont{12}{14.4}{\rmdefault}{\mddefault}{\updefault}{\color[rgb]{0,0,0}$q^{Db}_n$}%
}}}}
\put(1501,-136){\makebox(0,0)[lb]{\smash{{\SetFigFont{12}{14.4}{\rmdefault}{\mddefault}{\updefault}{\color[rgb]{0,0,0}$=-ig_3\g^\mu T^c_{ba} P_L$}%
}}}}
\put(1351,-736){\makebox(0,0)[lb]{\smash{{\SetFigFont{12}{14.4}{\rmdefault}{\mddefault}{\updefault}{\color[rgb]{0,0,0}$\bar{q}^a_0$}%
}}}}
\put(  1,-61){\makebox(0,0)[lb]{\smash{{\SetFigFont{12}{14.4}{\rmdefault}{\mddefault}{\updefault}{\color[rgb]{0,0,0}$G^c_n$}%
}}}}
\end{picture}%

%% file: G1-qs1-q0.pstex_t
\begin{picture}(0,0)%
\includegraphics{G1-qs1-q0.pstex}%
\end{picture}%
\setlength{\unitlength}{3947sp}%
\begingroup\makeatletter\ifx\SetFigFont\undefined%
\gdef\SetFigFont#1#2#3#4#5{%
  \reset@font\fontsize{#1}{#2pt}%
  \fontfamily{#3}\fontseries{#4}\fontshape{#5}%
  \selectfont}%
\fi\endgroup%
\begin{picture}(1530,1669)(-14,-809)
\put(1351,689){\makebox(0,0)[lb]{\smash{{\SetFigFont{12}{14.4}{\rmdefault}{\mddefault}{\updefault}{\color[rgb]{0,0,0}$q^{Sb}_n$}%
}}}}
\put(1501,-136){\makebox(0,0)[lb]{\smash{{\SetFigFont{12}{14.4}{\rmdefault}{\mddefault}{\updefault}{\color[rgb]{0,0,0}$=-ig_3\g^\mu T^c_{ba} P_R$}%
}}}}
\put(1351,-736){\makebox(0,0)[lb]{\smash{{\SetFigFont{12}{14.4}{\rmdefault}{\mddefault}{\updefault}{\color[rgb]{0,0,0}$\bar{q}^a_0$}%
}}}}
\put(  1,-61){\makebox(0,0)[lb]{\smash{{\SetFigFont{12}{14.4}{\rmdefault}{\mddefault}{\updefault}{\color[rgb]{0,0,0}$G^c_n$}%
}}}}
\end{picture}%

%% file: G-G1-G1.pstex_t
\begin{picture}(0,0)%
\includegraphics{G-G1-G1.pstex}%
\end{picture}%
\setlength{\unitlength}{3947sp}%
\begingroup\makeatletter\ifx\SetFigFont\undefined%
\gdef\SetFigFont#1#2#3#4#5{%
  \reset@font\fontsize{#1}{#2pt}%
  \fontfamily{#3}\fontseries{#4}\fontshape{#5}%
  \selectfont}%
\fi\endgroup%
\begin{picture}(1530,1503)(-89,-805)
\put(1276,-736){\makebox(0,0)[lb]{\smash{{\SetFigFont{12}{14.4}{\rmdefault}{\mddefault}{\updefault}{\color[rgb]{0,0,0}$G^{\lambda c}_n$}%
}}}}
\put(1426,-61){\makebox(0,0)[lb]{\smash{{\SetFigFont{12}{14.4}{\rmdefault}{\mddefault}{\updefault}{\color[rgb]{0,0,0}$=ig_3 f^{abc}\left [ (p-q)_\lambda g_{\mu\nu}+(q-r)_\mu g_{\lambda\nu}+(r-p)_\nu g_{\lambda\mu} \right ]$}%
}}}}
\put(901,464){\makebox(0,0)[lb]{\smash{{\SetFigFont{12}{14.4}{\rmdefault}{\mddefault}{\updefault}{\color[rgb]{0,0,0}$p$}%
}}}}
\put(526,-436){\makebox(0,0)[lb]{\smash{{\SetFigFont{12}{14.4}{\rmdefault}{\mddefault}{\updefault}{\color[rgb]{0,0,0}$q$}%
}}}}
\put(1426,-361){\makebox(0,0)[lb]{\smash{{\SetFigFont{12}{14.4}{\rmdefault}{\mddefault}{\updefault}{\color[rgb]{0,0,0}$r$}%
}}}}
\put(-74,-61){\makebox(0,0)[lb]{\smash{{\SetFigFont{12}{14.4}{\rmdefault}{\mddefault}{\updefault}{\color[rgb]{0,0,0}$G^{\nu b}$}%
}}}}
\put(1276,539){\makebox(0,0)[lb]{\smash{{\SetFigFont{12}{14.4}{\rmdefault}{\mddefault}{\updefault}{\color[rgb]{0,0,0}$G^{\mu a}_n$}%
}}}}
\end{picture}%

%% file: G-G-G1-G1.pstex_t
\begin{picture}(0,0)%
\includegraphics{G-G-G1-G1.pstex}%
\end{picture}%
\setlength{\unitlength}{3947sp}%
\begingroup\makeatletter\ifx\SetFigFont\undefined%
\gdef\SetFigFont#1#2#3#4#5{%
  \reset@font\fontsize{#1}{#2pt}%
  \fontfamily{#3}\fontseries{#4}\fontshape{#5}%
  \selectfont}%
\fi\endgroup%
\begin{picture}(2355,1578)(61,-880)
\put(1426,-811){\makebox(0,0)[lb]{\smash{{\SetFigFont{12}{14.4}{\rmdefault}{\mddefault}{\updefault}{\color[rgb]{0,0,0}$G_n^{\rho d}$}%
}}}}
\put( 76,539){\makebox(0,0)[lb]{\smash{{\SetFigFont{12}{14.4}{\rmdefault}{\mddefault}{\updefault}{\color[rgb]{0,0,0}$G^{\nu b}$}%
}}}}
\put( 76,-811){\makebox(0,0)[lb]{\smash{{\SetFigFont{12}{14.4}{\rmdefault}{\mddefault}{\updefault}{\color[rgb]{0,0,0}$G^{\mu a}$}%
}}}}
\put(1501, 89){\makebox(0,0)[lb]{\smash{{\SetFigFont{12}{14.4}{\rmdefault}{\mddefault}{\updefault}{\color[rgb]{0,0,0}$=-i g_3^2 \left [ f^{abe}f^{cde}(g_{\lambda\nu}g_{\mu\rho}-g_{\lambda\rho}g_{\mu\nu}) + f^{acd}f^{bde}(g_{\lambda\mu}g_{\nu\rho}-g_{\lambda\rho}g_{\mu\nu}) \right . $}%
}}}}
\put(2401,-361){\makebox(0,0)[lb]{\smash{{\SetFigFont{12}{14.4}{\rmdefault}{\mddefault}{\updefault}{\color[rgb]{0,0,0}$+ \left .f^{ade}f^{bce}(g_{\lambda\mu}g_{\nu\rho}-g_{\lambda\nu}g_{\mu\rho}) \right ] $}%
}}}}
\put(1426,539){\makebox(0,0)[lb]{\smash{{\SetFigFont{12}{14.4}{\rmdefault}{\mddefault}{\updefault}{\color[rgb]{0,0,0}$G_n^{\lambda c}$}%
}}}}
\end{picture}%

%% file: G1-G1-G1-G1.pstex_t
\begin{picture}(0,0)%
\includegraphics{G1-G1-G1-G1.pstex}%
\end{picture}%
\setlength{\unitlength}{3947sp}%
\begingroup\makeatletter\ifx\SetFigFont\undefined%
\gdef\SetFigFont#1#2#3#4#5{%
  \reset@font\fontsize{#1}{#2pt}%
  \fontfamily{#3}\fontseries{#4}\fontshape{#5}%
  \selectfont}%
\fi\endgroup%
\begin{picture}(2355,1578)(61,-880)
\put(1426,-811){\makebox(0,0)[lb]{\smash{{\SetFigFont{12}{14.4}{\rmdefault}{\mddefault}{\updefault}{\color[rgb]{0,0,0}$G_n^{\rho d}$}%
}}}}
\put(1501, 89){\makebox(0,0)[lb]{\smash{{\SetFigFont{12}{14.4}{\rmdefault}{\mddefault}{\updefault}{\color[rgb]{0,0,0}$=-i\frac{3}{2}g_3^2 \left [ f^{abe}f^{cde}(g_{\lambda\nu}g_{\mu\rho}-g_{\lambda\rho}g_{\mu\nu}) + f^{acd}f^{bde}(g_{\lambda\mu}g_{\nu\rho}-g_{\lambda\rho}g_{\mu\nu}) \right . $}%
}}}}
\put(2401,-361){\makebox(0,0)[lb]{\smash{{\SetFigFont{12}{14.4}{\rmdefault}{\mddefault}{\updefault}{\color[rgb]{0,0,0}$+ \left .f^{ade}f^{bce}(g_{\lambda\mu}g_{\nu\rho}-g_{\lambda\nu}g_{\mu\rho}) \right ] $}%
}}}}
\put( 76,539){\makebox(0,0)[lb]{\smash{{\SetFigFont{12}{14.4}{\rmdefault}{\mddefault}{\updefault}{\color[rgb]{0,0,0}$G_n^{\nu b}$}%
}}}}
\put(1426,539){\makebox(0,0)[lb]{\smash{{\SetFigFont{12}{14.4}{\rmdefault}{\mddefault}{\updefault}{\color[rgb]{0,0,0}$G_n^{\lambda c}$}%
}}}}
\put( 76,-811){\makebox(0,0)[lb]{\smash{{\SetFigFont{12}{14.4}{\rmdefault}{\mddefault}{\updefault}{\color[rgb]{0,0,0}$G_n^{\mu a}$}%
}}}}
\end{picture}%

%% file: A-f1-f1.pstex_t
\begin{picture}(0,0)%
\includegraphics{A-f1-f1.pstex}%
\end{picture}%
\setlength{\unitlength}{3947sp}%
\begingroup\makeatletter\ifx\SetFigFont\undefined%
\gdef\SetFigFont#1#2#3#4#5{%
  \reset@font\fontsize{#1}{#2pt}%
  \fontfamily{#3}\fontseries{#4}\fontshape{#5}%
  \selectfont}%
\fi\endgroup%
\begin{picture}(1455,1653)(61,-805)
\put(1426,-736){\makebox(0,0)[lb]{\smash{{\SetFigFont{12}{14.4}{\rmdefault}{\mddefault}{\updefault}{\color[rgb]{0,0,0}$\bar{f}_n$}%
}}}}
\put(1501,-61){\makebox(0,0)[lb]{\smash{{\SetFigFont{12}{14.4}{\rmdefault}{\mddefault}{\updefault}{\color[rgb]{0,0,0}$=-iQ_f e \g^\mu$}%
}}}}
\put( 76, 14){\makebox(0,0)[lb]{\smash{{\SetFigFont{12}{14.4}{\rmdefault}{\mddefault}{\updefault}{\color[rgb]{0,0,0}$\g$}%
}}}}
\put(1426,689){\makebox(0,0)[lb]{\smash{{\SetFigFont{12}{14.4}{\rmdefault}{\mddefault}{\updefault}{\color[rgb]{0,0,0}$f_n$}%
}}}}
\end{picture}%

%% file: Z-fd1-fd1.pstex_t
\begin{picture}(0,0)%
\includegraphics{Z-fd1-fd1.pstex}%
\end{picture}%
\setlength{\unitlength}{3947sp}%
\begingroup\makeatletter\ifx\SetFigFont\undefined%
\gdef\SetFigFont#1#2#3#4#5{%
  \reset@font\fontsize{#1}{#2pt}%
  \fontfamily{#3}\fontseries{#4}\fontshape{#5}%
  \selectfont}%
\fi\endgroup%
\begin{picture}(1455,1653)(61,-805)
\put(1426,689){\makebox(0,0)[lb]{\smash{{\SetFigFont{12}{14.4}{\rmdefault}{\mddefault}{\updefault}{\color[rgb]{0,0,0}$f^D_n$}%
}}}}
\put( 76, 14){\makebox(0,0)[lb]{\smash{{\SetFigFont{12}{14.4}{\rmdefault}{\mddefault}{\updefault}{\color[rgb]{0,0,0}$Z$}%
}}}}
\put(1501, 14){\makebox(0,0)[lb]{\smash{{\SetFigFont{12}{14.4}{\rmdefault}{\mddefault}{\updefault}{\color[rgb]{0,0,0}$=-i\frac{g_2}{\cos\theta_W}c_L\g^\mu$}%
}}}}
\put(1426,-736){\makebox(0,0)[lb]{\smash{{\SetFigFont{12}{14.4}{\rmdefault}{\mddefault}{\updefault}{\color[rgb]{0,0,0}$\bar{f}^D_n$}%
}}}}
\end{picture}%

%% file: Z-fs1-fs1.pstex_t
\begin{picture}(0,0)%
\includegraphics{Z-fs1-fs1.pstex}%
\end{picture}%
\setlength{\unitlength}{3947sp}%
\begingroup\makeatletter\ifx\SetFigFont\undefined%
\gdef\SetFigFont#1#2#3#4#5{%
  \reset@font\fontsize{#1}{#2pt}%
  \fontfamily{#3}\fontseries{#4}\fontshape{#5}%
  \selectfont}%
\fi\endgroup%
\begin{picture}(1455,1653)(61,-805)
\put(1426,-736){\makebox(0,0)[lb]{\smash{{\SetFigFont{12}{14.4}{\rmdefault}{\mddefault}{\updefault}{\color[rgb]{0,0,0}$\bar{f}^S_n$}%
}}}}
\put( 76, 14){\makebox(0,0)[lb]{\smash{{\SetFigFont{12}{14.4}{\rmdefault}{\mddefault}{\updefault}{\color[rgb]{0,0,0}$Z$}%
}}}}
\put(1501, 14){\makebox(0,0)[lb]{\smash{{\SetFigFont{12}{14.4}{\rmdefault}{\mddefault}{\updefault}{\color[rgb]{0,0,0}$=-i\frac{g_2}{\cos\theta_W}c_R\g^\mu$}%
}}}}
\put(1426,689){\makebox(0,0)[lb]{\smash{{\SetFigFont{12}{14.4}{\rmdefault}{\mddefault}{\updefault}{\color[rgb]{0,0,0}$f^S_n$}%
}}}}
\end{picture}%

%% file: W-fd1-fd1.pstex_t
\begin{picture}(0,0)%
\includegraphics{W-fd1-fd1.pstex}%
\end{picture}%
\setlength{\unitlength}{3947sp}%
\begingroup\makeatletter\ifx\SetFigFont\undefined%
\gdef\SetFigFont#1#2#3#4#5{%
  \reset@font\fontsize{#1}{#2pt}%
  \fontfamily{#3}\fontseries{#4}\fontshape{#5}%
  \selectfont}%
\fi\endgroup%
\begin{picture}(1530,1653)(-14,-805)
\put(1426,-736){\makebox(0,0)[lb]{\smash{{\SetFigFont{12}{14.4}{\rmdefault}{\mddefault}{\updefault}{\color[rgb]{0,0,0}$\bar{f'}^D_n$}%
}}}}
\put(1501, 14){\makebox(0,0)[lb]{\smash{{\SetFigFont{12}{14.4}{\rmdefault}{\mddefault}{\updefault}{\color[rgb]{0,0,0}$=-i\frac{g_2}{\sqrt{2}} \g^\mu V_{ff'}$}%
}}}}
\put(  1, 14){\makebox(0,0)[lb]{\smash{{\SetFigFont{12}{14.4}{\rmdefault}{\mddefault}{\updefault}{\color[rgb]{0,0,0}$W^\pm$}%
}}}}
\put(1426,689){\makebox(0,0)[lb]{\smash{{\SetFigFont{12}{14.4}{\rmdefault}{\mddefault}{\updefault}{\color[rgb]{0,0,0}$f^D_n$}%
}}}}
\end{picture}%

%% file: B1-fd1-fd1.pstex_t
\begin{picture}(0,0)%
\includegraphics{B1-fd1-fd1.pstex}%
\end{picture}%
\setlength{\unitlength}{3947sp}%
\begingroup\makeatletter\ifx\SetFigFont\undefined%
\gdef\SetFigFont#1#2#3#4#5{%
  \reset@font\fontsize{#1}{#2pt}%
  \fontfamily{#3}\fontseries{#4}\fontshape{#5}%
  \selectfont}%
\fi\endgroup%
\begin{picture}(1530,1653)(-14,-805)
\put(1426,-736){\makebox(0,0)[lb]{\smash{{\SetFigFont{12}{14.4}{\rmdefault}{\mddefault}{\updefault}{\color[rgb]{0,0,0}$\bar{f}^D_n$}%
}}}}
\put(1501, 14){\makebox(0,0)[lb]{\smash{{\SetFigFont{12}{14.4}{\rmdefault}{\mddefault}{\updefault}{\color[rgb]{0,0,0}$=-i \frac{Y}{2}g_1 \g^\mu P_L$}%
}}}}
\put(1426,689){\makebox(0,0)[lb]{\smash{{\SetFigFont{12}{14.4}{\rmdefault}{\mddefault}{\updefault}{\color[rgb]{0,0,0}$f^D_0$}%
}}}}
\put(  1, 14){\makebox(0,0)[lb]{\smash{{\SetFigFont{12}{14.4}{\rmdefault}{\mddefault}{\updefault}{\color[rgb]{0,0,0}$B_n$}%
}}}}
\end{picture}%

%% file: B1-fs1-fs1.pstex_t
\begin{picture}(0,0)%
\includegraphics{B1-fs1-fs1.pstex}%
\end{picture}%
\setlength{\unitlength}{3947sp}%
\begingroup\makeatletter\ifx\SetFigFont\undefined%
\gdef\SetFigFont#1#2#3#4#5{%
  \reset@font\fontsize{#1}{#2pt}%
  \fontfamily{#3}\fontseries{#4}\fontshape{#5}%
  \selectfont}%
\fi\endgroup%
\begin{picture}(1530,1653)(-14,-805)
\put(  1, 14){\makebox(0,0)[lb]{\smash{{\SetFigFont{12}{14.4}{\rmdefault}{\mddefault}{\updefault}{\color[rgb]{0,0,0}$B_n$}%
}}}}
\put(1501, 14){\makebox(0,0)[lb]{\smash{{\SetFigFont{12}{14.4}{\rmdefault}{\mddefault}{\updefault}{\color[rgb]{0,0,0}$=-i\frac{Y}{2}g_1 \g^\mu P_R$}%
}}}}
\put(1426,689){\makebox(0,0)[lb]{\smash{{\SetFigFont{12}{14.4}{\rmdefault}{\mddefault}{\updefault}{\color[rgb]{0,0,0}$f^S_0$}%
}}}}
\put(1426,-736){\makebox(0,0)[lb]{\smash{{\SetFigFont{12}{14.4}{\rmdefault}{\mddefault}{\updefault}{\color[rgb]{0,0,0}$\bar{f}^S_n$}%
}}}}
\end{picture}%

%% file: Z1-fd1-f0.pstex_t
\begin{picture}(0,0)%
\includegraphics{Z1-fd1-f0.pstex}%
\end{picture}%
\setlength{\unitlength}{3947sp}%
\begingroup\makeatletter\ifx\SetFigFont\undefined%
\gdef\SetFigFont#1#2#3#4#5{%
  \reset@font\fontsize{#1}{#2pt}%
  \fontfamily{#3}\fontseries{#4}\fontshape{#5}%
  \selectfont}%
\fi\endgroup%
\begin{picture}(1530,1653)(-14,-805)
\put(1426,689){\makebox(0,0)[lb]{\smash{{\SetFigFont{12}{14.4}{\rmdefault}{\mddefault}{\updefault}{\color[rgb]{0,0,0}$f^D_n$}%
}}}}
\put(1426,-736){\makebox(0,0)[lb]{\smash{{\SetFigFont{12}{14.4}{\rmdefault}{\mddefault}{\updefault}{\color[rgb]{0,0,0}$\bar{f}_0$}%
}}}}
\put(1501, 14){\makebox(0,0)[lb]{\smash{{\SetFigFont{12}{14.4}{\rmdefault}{\mddefault}{\updefault}{\color[rgb]{0,0,0}$=-i I_3 g_2 \g^\mu P_L$}%
}}}}
\put(  1, 14){\makebox(0,0)[lb]{\smash{{\SetFigFont{12}{14.4}{\rmdefault}{\mddefault}{\updefault}{\color[rgb]{0,0,0}$Z_n$}%
}}}}
\end{picture}%

%% file: W1-fd1-f0.pstex_t
\begin{picture}(0,0)%
\includegraphics{W1-fd1-f0.pstex}%
\end{picture}%
\setlength{\unitlength}{3947sp}%
\begingroup\makeatletter\ifx\SetFigFont\undefined%
\gdef\SetFigFont#1#2#3#4#5{%
  \reset@font\fontsize{#1}{#2pt}%
  \fontfamily{#3}\fontseries{#4}\fontshape{#5}%
  \selectfont}%
\fi\endgroup%
\begin{picture}(1530,1728)(-14,-880)
\put(  1, 14){\makebox(0,0)[lb]{\smash{{\SetFigFont{12}{14.4}{\rmdefault}{\mddefault}{\updefault}{\color[rgb]{0,0,0}$W^\pm_n$}%
}}}}
\put(1501, 14){\makebox(0,0)[lb]{\smash{{\SetFigFont{12}{14.4}{\rmdefault}{\mddefault}{\updefault}{\color[rgb]{0,0,0}$=-i\frac{g_2}{\sqrt{2}} \g^\mu P_L V_{ff'}$}%
}}}}
\put(1426,-811){\makebox(0,0)[lb]{\smash{{\SetFigFont{12}{14.4}{\rmdefault}{\mddefault}{\updefault}{\color[rgb]{0,0,0}$\bar{f'}_0$}%
}}}}
\put(1426,689){\makebox(0,0)[lb]{\smash{{\SetFigFont{12}{14.4}{\rmdefault}{\mddefault}{\updefault}{\color[rgb]{0,0,0}$f^D_n$}%
}}}}
\end{picture}%

%% file: B2-fd1-fd1.pstex_t
\begin{picture}(0,0)%
\includegraphics{B2-fd1-fd1.pstex}%
\end{picture}%
\setlength{\unitlength}{3947sp}%
\begingroup\makeatletter\ifx\SetFigFont\undefined%
\gdef\SetFigFont#1#2#3#4#5{%
  \reset@font\fontsize{#1}{#2pt}%
  \fontfamily{#3}\fontseries{#4}\fontshape{#5}%
  \selectfont}%
\fi\endgroup%
\begin{picture}(1500,1639)(1,-794)
\put(  1, 14){\makebox(0,0)[lb]{\smash{\SetFigFont{12}{14.4}{\rmdefault}{\mddefault}{\updefault}{\color[rgb]{0,0,0}$B_2$}%
}}}
\put(1426,689){\makebox(0,0)[lb]{\smash{\SetFigFont{12}{14.4}{\rmdefault}{\mddefault}{\updefault}{\color[rgb]{0,0,0}$f^D_1$}%
}}}
\put(1501, 14){\makebox(0,0)[lb]{\smash{\SetFigFont{12}{14.4}{\rmdefault}{\mddefault}{\updefault}{\color[rgb]{0,0,0}$=i \frac{Y}{2} \frac{g_1}{\sqrt{2}} \g^\mu \g^5$}%
}}}
\put(1426,-736){\makebox(0,0)[lb]{\smash{\SetFigFont{12}{14.4}{\rmdefault}{\mddefault}{\updefault}{\color[rgb]{0,0,0}$\bar{f}^D_1$}%
}}}
\end{picture}

%% file: B2-fs1-fs1.pstex_t
\begin{picture}(0,0)%
\includegraphics{B2-fs1-fs1.pstex}%
\end{picture}%
\setlength{\unitlength}{3947sp}%
\begingroup\makeatletter\ifx\SetFigFont\undefined%
\gdef\SetFigFont#1#2#3#4#5{%
  \reset@font\fontsize{#1}{#2pt}%
  \fontfamily{#3}\fontseries{#4}\fontshape{#5}%
  \selectfont}%
\fi\endgroup%
\begin{picture}(1530,1653)(-14,-805)
\put(1501, 14){\makebox(0,0)[lb]{\smash{{\SetFigFont{12}{14.4}{\rmdefault}{\mddefault}{\updefault}{\color[rgb]{0,0,0}$=-i \frac{Y}{2}\frac{g_1}{\sqrt{2}} \g^\mu \g^5$}%
}}}}
\put(  1, 14){\makebox(0,0)[lb]{\smash{{\SetFigFont{12}{14.4}{\rmdefault}{\mddefault}{\updefault}{\color[rgb]{0,0,0}$B_2$}%
}}}}
\put(1426,689){\makebox(0,0)[lb]{\smash{{\SetFigFont{12}{14.4}{\rmdefault}{\mddefault}{\updefault}{\color[rgb]{0,0,0}$f^S_1$}%
}}}}
\put(1426,-736){\makebox(0,0)[lb]{\smash{{\SetFigFont{12}{14.4}{\rmdefault}{\mddefault}{\updefault}{\color[rgb]{0,0,0}$\bar{f}^S_1$}%
}}}}
\end{picture}%

%% file: Z2-fd1-fd1.pstex_t
\begin{picture}(0,0)%
\includegraphics{Z2-fd1-fd1.pstex}%
\end{picture}%
\setlength{\unitlength}{3947sp}%
\begingroup\makeatletter\ifx\SetFigFont\undefined%
\gdef\SetFigFont#1#2#3#4#5{%
  \reset@font\fontsize{#1}{#2pt}%
  \fontfamily{#3}\fontseries{#4}\fontshape{#5}%
  \selectfont}%
\fi\endgroup%
\begin{picture}(1530,1653)(-14,-805)
\put(1426,-736){\makebox(0,0)[lb]{\smash{{\SetFigFont{12}{14.4}{\rmdefault}{\mddefault}{\updefault}{\color[rgb]{0,0,0}$\bar{f}_1^D$}%
}}}}
\put(  1, 14){\makebox(0,0)[lb]{\smash{{\SetFigFont{12}{14.4}{\rmdefault}{\mddefault}{\updefault}{\color[rgb]{0,0,0}$Z_2$}%
}}}}
\put(1501, 14){\makebox(0,0)[lb]{\smash{{\SetFigFont{12}{14.4}{\rmdefault}{\mddefault}{\updefault}{\color[rgb]{0,0,0}$=i I_3\frac{g_2}{\sqrt{2}} \g^\mu \g^5$}%
}}}}
\put(1426,689){\makebox(0,0)[lb]{\smash{{\SetFigFont{12}{14.4}{\rmdefault}{\mddefault}{\updefault}{\color[rgb]{0,0,0}$f_1^D$}%
}}}}
\end{picture}%

%% file: W2-fd1-fd1.pstex_t
\begin{picture}(0,0)%
\includegraphics{W2-fd1-fd1.pstex}%
\end{picture}%
\setlength{\unitlength}{3947sp}%
\begingroup\makeatletter\ifx\SetFigFont\undefined%
\gdef\SetFigFont#1#2#3#4#5{%
  \reset@font\fontsize{#1}{#2pt}%
  \fontfamily{#3}\fontseries{#4}\fontshape{#5}%
  \selectfont}%
\fi\endgroup%
\begin{picture}(1530,1728)(-14,-880)
\put(1426,-811){\makebox(0,0)[lb]{\smash{{\SetFigFont{12}{14.4}{\rmdefault}{\mddefault}{\updefault}{\color[rgb]{0,0,0}$\bar{f'}_1^D$}%
}}}}
\put(  1, 14){\makebox(0,0)[lb]{\smash{{\SetFigFont{12}{14.4}{\rmdefault}{\mddefault}{\updefault}{\color[rgb]{0,0,0}$W^\pm_2$}%
}}}}
\put(1426,689){\makebox(0,0)[lb]{\smash{{\SetFigFont{12}{14.4}{\rmdefault}{\mddefault}{\updefault}{\color[rgb]{0,0,0}$f^D_1$}%
}}}}
\put(1501, 14){\makebox(0,0)[lb]{\smash{{\SetFigFont{12}{14.4}{\rmdefault}{\mddefault}{\updefault}{\color[rgb]{0,0,0}$=-i\frac{g_2}{2} \g^\mu P_L V_{ff'}$}%
}}}}
\end{picture}%

%% file: A-W1-W1.pstex_t
\begin{picture}(0,0)%
\includegraphics{A-W1-W1.pstex}%
\end{picture}%
\setlength{\unitlength}{3947sp}%
\begingroup\makeatletter\ifx\SetFigFont\undefined%
\gdef\SetFigFont#1#2#3#4#5{%
  \reset@font\fontsize{#1}{#2pt}%
  \fontfamily{#3}\fontseries{#4}\fontshape{#5}%
  \selectfont}%
\fi\endgroup%
\begin{picture}(1830,1653)(-14,-805)
\put(1426,-736){\makebox(0,0)[lb]{\smash{{\SetFigFont{12}{14.4}{\rmdefault}{\mddefault}{\updefault}{\color[rgb]{0,0,0}$W^{n-}_{\lambda}$}%
}}}}
\put(601,-361){\makebox(0,0)[lb]{\smash{{\SetFigFont{12}{14.4}{\rmdefault}{\mddefault}{\updefault}{\color[rgb]{0,0,0}$k_1$}%
}}}}
\put(901,389){\makebox(0,0)[lb]{\smash{{\SetFigFont{12}{14.4}{\rmdefault}{\mddefault}{\updefault}{\color[rgb]{0,0,0}$k_2$}%
}}}}
\put(1801,-61){\makebox(0,0)[lb]{\smash{{\SetFigFont{12}{14.4}{\rmdefault}{\mddefault}{\updefault}{\color[rgb]{0,0,0}$=-ie [ (k_1 - k_2 )g_{\mu\nu}+(k_2 -k_3 )g_{\nu\lambda}+(k_3 - k_1 )g_{\lambda\mu} ]$}%
}}}}
\put(1426,-136){\makebox(0,0)[lb]{\smash{{\SetFigFont{12}{14.4}{\rmdefault}{\mddefault}{\updefault}{\color[rgb]{0,0,0}$k_3$}%
}}}}
\put(  1, 14){\makebox(0,0)[lb]{\smash{{\SetFigFont{12}{14.4}{\rmdefault}{\mddefault}{\updefault}{\color[rgb]{0,0,0}$A_\mu$}%
}}}}
\put(1426,689){\makebox(0,0)[lb]{\smash{{\SetFigFont{12}{14.4}{\rmdefault}{\mddefault}{\updefault}{\color[rgb]{0,0,0}$W^{n+}_\nu$}%
}}}}
\end{picture}%

%% file: Z1-W-W1.pstex_t
\begin{picture}(0,0)%
\includegraphics{Z1-W-W1.pstex}%
\end{picture}%
\setlength{\unitlength}{3947sp}%
\begingroup\makeatletter\ifx\SetFigFont\undefined%
\gdef\SetFigFont#1#2#3#4#5{%
  \reset@font\fontsize{#1}{#2pt}%
  \fontfamily{#3}\fontseries{#4}\fontshape{#5}%
  \selectfont}%
\fi\endgroup%
\begin{picture}(1830,1653)(-14,-805)
\put(1426,-736){\makebox(0,0)[lb]{\smash{{\SetFigFont{12}{14.4}{\rmdefault}{\mddefault}{\updefault}{\color[rgb]{0,0,0}$W^{n-}_{\lambda}$}%
}}}}
\put(1426,689){\makebox(0,0)[lb]{\smash{{\SetFigFont{12}{14.4}{\rmdefault}{\mddefault}{\updefault}{\color[rgb]{0,0,0}$W^{+}_\nu$}%
}}}}
\put(601,-361){\makebox(0,0)[lb]{\smash{{\SetFigFont{12}{14.4}{\rmdefault}{\mddefault}{\updefault}{\color[rgb]{0,0,0}$k_1$}%
}}}}
\put(901,389){\makebox(0,0)[lb]{\smash{{\SetFigFont{12}{14.4}{\rmdefault}{\mddefault}{\updefault}{\color[rgb]{0,0,0}$k_2$}%
}}}}
\put(1426,-136){\makebox(0,0)[lb]{\smash{{\SetFigFont{12}{14.4}{\rmdefault}{\mddefault}{\updefault}{\color[rgb]{0,0,0}$k_3$}%
}}}}
\put(1801,-61){\makebox(0,0)[lb]{\smash{{\SetFigFont{12}{14.4}{\rmdefault}{\mddefault}{\updefault}{\color[rgb]{0,0,0}$=-ig_2 [ (k_1 - k_2 )g_{\mu\nu}+(k_2 -k_3 )g_{\nu\lambda}+(k_3 - k_1 )g_{\lambda\mu} ]$}%
}}}}
\put(  1, 14){\makebox(0,0)[lb]{\smash{{\SetFigFont{12}{14.4}{\rmdefault}{\mddefault}{\updefault}{\color[rgb]{0,0,0}$Z^n_\mu$}%
}}}}
\end{picture}%

%% file: Z-W1-W1.pstex_t
\begin{picture}(0,0)%
\includegraphics{Z-W1-W1.pstex}%
\end{picture}%
\setlength{\unitlength}{3947sp}%
\begingroup\makeatletter\ifx\SetFigFont\undefined%
\gdef\SetFigFont#1#2#3#4#5{%
  \reset@font\fontsize{#1}{#2pt}%
  \fontfamily{#3}\fontseries{#4}\fontshape{#5}%
  \selectfont}%
\fi\endgroup%
\begin{picture}(1830,1653)(-14,-805)
\put(1426,-736){\makebox(0,0)[lb]{\smash{{\SetFigFont{12}{14.4}{\rmdefault}{\mddefault}{\updefault}{\color[rgb]{0,0,0}$W^{n-}_{\lambda}$}%
}}}}
\put(  1, 14){\makebox(0,0)[lb]{\smash{{\SetFigFont{12}{14.4}{\rmdefault}{\mddefault}{\updefault}{\color[rgb]{0,0,0}$Z_\mu$}%
}}}}
\put(601,-361){\makebox(0,0)[lb]{\smash{{\SetFigFont{12}{14.4}{\rmdefault}{\mddefault}{\updefault}{\color[rgb]{0,0,0}$k_1$}%
}}}}
\put(901,389){\makebox(0,0)[lb]{\smash{{\SetFigFont{12}{14.4}{\rmdefault}{\mddefault}{\updefault}{\color[rgb]{0,0,0}$k_2$}%
}}}}
\put(1426,-136){\makebox(0,0)[lb]{\smash{{\SetFigFont{12}{14.4}{\rmdefault}{\mddefault}{\updefault}{\color[rgb]{0,0,0}$k_3$}%
}}}}
\put(1801,-61){\makebox(0,0)[lb]{\smash{{\SetFigFont{12}{14.4}{\rmdefault}{\mddefault}{\updefault}{\color[rgb]{0,0,0}$=-ig_2 \cos\theta_W [ (k_1 - k_2 )g_{\mu\nu}+(k_2 -k_3 )g_{\nu\lambda}+(k_3 - k_1 )g_{\lambda\mu} ]$}%
}}}}
\put(1426,689){\makebox(0,0)[lb]{\smash{{\SetFigFont{12}{14.4}{\rmdefault}{\mddefault}{\updefault}{\color[rgb]{0,0,0}$W^{n+}_\nu$}%
}}}}
\end{picture}%

%% file: Z2-W1-W1.pstex_t
\begin{picture}(0,0)%
\includegraphics{Z2-W1-W1.pstex}%
\end{picture}%
\setlength{\unitlength}{3947sp}%
\begingroup\makeatletter\ifx\SetFigFont\undefined%
\gdef\SetFigFont#1#2#3#4#5{%
  \reset@font\fontsize{#1}{#2pt}%
  \fontfamily{#3}\fontseries{#4}\fontshape{#5}%
  \selectfont}%
\fi\endgroup%
\begin{picture}(1800,1639)(1,-794)
\put(1426,689){\makebox(0,0)[lb]{\smash{\SetFigFont{12}{14.4}{\rmdefault}{\mddefault}{\updefault}{\color[rgb]{0,0,0}$W^{1+}_\nu$}%
}}}
\put(1426,-736){\makebox(0,0)[lb]{\smash{\SetFigFont{12}{14.4}{\rmdefault}{\mddefault}{\updefault}{\color[rgb]{0,0,0}$W^{1-}_{\lambda}$}%
}}}
\put(  1, 14){\makebox(0,0)[lb]{\smash{\SetFigFont{12}{14.4}{\rmdefault}{\mddefault}{\updefault}{\color[rgb]{0,0,0}$Z^2_\mu$}%
}}}
\put(601,-361){\makebox(0,0)[lb]{\smash{\SetFigFont{12}{14.4}{\rmdefault}{\mddefault}{\updefault}{\color[rgb]{0,0,0}$k_1$}%
}}}
\put(901,389){\makebox(0,0)[lb]{\smash{\SetFigFont{12}{14.4}{\rmdefault}{\mddefault}{\updefault}{\color[rgb]{0,0,0}$k_2$}%
}}}
\put(1426,-136){\makebox(0,0)[lb]{\smash{\SetFigFont{12}{14.4}{\rmdefault}{\mddefault}{\updefault}{\color[rgb]{0,0,0}$k_3$}%
}}}
\put(1801,-61){\makebox(0,0)[lb]{\smash{\SetFigFont{12}{14.4}{\rmdefault}{\mddefault}{\updefault}{\color[rgb]{0,0,0}$=-i\frac{g_2}{\sqrt 2} \cos\theta_W [ (k_1 - k_2 )g_{\mu\nu}+(k_2 -k_3 )g_{\nu\lambda}+(k_3 - k_1 )g_{\lambda\mu} ]$}%
}}}
\end{picture}

%% file: A-A-W1-W1.pstex_t
\begin{picture}(0,0)%
\includegraphics{A-A-W1-W1.pstex}%
\end{picture}%
\setlength{\unitlength}{3947sp}%
\begingroup\makeatletter\ifx\SetFigFont\undefined%
\gdef\SetFigFont#1#2#3#4#5{%
  \reset@font\fontsize{#1}{#2pt}%
  \fontfamily{#3}\fontseries{#4}\fontshape{#5}%
  \selectfont}%
\fi\endgroup%
\begin{picture}(1755,1657)(136,-809)
\put(1276,-736){\makebox(0,0)[lb]{\smash{{\SetFigFont{12}{14.4}{\rmdefault}{\mddefault}{\updefault}{\color[rgb]{0,0,0}$W^{n-}_{\sigma}$}%
}}}}
\put(151,689){\makebox(0,0)[lb]{\smash{{\SetFigFont{12}{14.4}{\rmdefault}{\mddefault}{\updefault}{\color[rgb]{0,0,0}$A_\mu$}%
}}}}
\put(151,-736){\makebox(0,0)[lb]{\smash{{\SetFigFont{12}{14.4}{\rmdefault}{\mddefault}{\updefault}{\color[rgb]{0,0,0}$A_\nu$}%
}}}}
\put(1876,-61){\makebox(0,0)[lb]{\smash{{\SetFigFont{12}{14.4}{\rmdefault}{\mddefault}{\updefault}{\color[rgb]{0,0,0}$=-ie^2 (2g^{\mu\nu}g^{\rho\sigma}-g^{\mu\rho}g^{\nu\sigma}-g^{\mu\sigma}g^{\nu\rho})$}%
}}}}
\put(1276,689){\makebox(0,0)[lb]{\smash{{\SetFigFont{12}{14.4}{\rmdefault}{\mddefault}{\updefault}{\color[rgb]{0,0,0}$W^{n+}_\rho$}%
}}}}
\end{picture}%

%% file: A-W-W1-Z1.pstex_t
\begin{picture}(0,0)%
\includegraphics{A-W-W1-Z1.pstex}%
\end{picture}%
\setlength{\unitlength}{3947sp}%
\begingroup\makeatletter\ifx\SetFigFont\undefined%
\gdef\SetFigFont#1#2#3#4#5{%
  \reset@font\fontsize{#1}{#2pt}%
  \fontfamily{#3}\fontseries{#4}\fontshape{#5}%
  \selectfont}%
\fi\endgroup%
\begin{picture}(1755,1657)(136,-809)
\put(1276,689){\makebox(0,0)[lb]{\smash{{\SetFigFont{12}{14.4}{\rmdefault}{\mddefault}{\updefault}{\color[rgb]{0,0,0}$W^{n-}_\rho$}%
}}}}
\put(151,689){\makebox(0,0)[lb]{\smash{{\SetFigFont{12}{14.4}{\rmdefault}{\mddefault}{\updefault}{\color[rgb]{0,0,0}$A_\mu$}%
}}}}
\put(1876,-61){\makebox(0,0)[lb]{\smash{{\SetFigFont{12}{14.4}{\rmdefault}{\mddefault}{\updefault}{\color[rgb]{0,0,0}$=-i\frac{e^2}{\sin\theta_W} (2g^{\mu\nu}g^{\rho\sigma}-g^{\mu\rho}g^{\nu\sigma}-g^{\mu\sigma}g^{\nu\rho})$}%
}}}}
\put(1276,-736){\makebox(0,0)[lb]{\smash{{\SetFigFont{12}{14.4}{\rmdefault}{\mddefault}{\updefault}{\color[rgb]{0,0,0}$W^{+}_\sigma$}%
}}}}
\put(151,-736){\makebox(0,0)[lb]{\smash{{\SetFigFont{12}{14.4}{\rmdefault}{\mddefault}{\updefault}{\color[rgb]{0,0,0}$Z^{n}_{\nu}$}%
}}}}
\end{picture}%

%% file: A-W1-W1-Z.pstex_t
\begin{picture}(0,0)%
\includegraphics{A-W1-W1-Z.pstex}%
\end{picture}%
\setlength{\unitlength}{3947sp}%
\begingroup\makeatletter\ifx\SetFigFont\undefined%
\gdef\SetFigFont#1#2#3#4#5{%
  \reset@font\fontsize{#1}{#2pt}%
  \fontfamily{#3}\fontseries{#4}\fontshape{#5}%
  \selectfont}%
\fi\endgroup%
\begin{picture}(1755,1657)(136,-809)
\put(1276,-736){\makebox(0,0)[lb]{\smash{{\SetFigFont{12}{14.4}{\rmdefault}{\mddefault}{\updefault}{\color[rgb]{0,0,0}$W^{n+}_\sigma$}%
}}}}
\put(151,689){\makebox(0,0)[lb]{\smash{{\SetFigFont{12}{14.4}{\rmdefault}{\mddefault}{\updefault}{\color[rgb]{0,0,0}$A_\mu$}%
}}}}
\put(1876,-61){\makebox(0,0)[lb]{\smash{{\SetFigFont{12}{14.4}{\rmdefault}{\mddefault}{\updefault}{\color[rgb]{0,0,0}$=-i\frac{\cos\theta_W e^2}{\sin\theta_W} (2g^{\mu\nu}g^{\rho\sigma}-g^{\mu\rho}g^{\nu\sigma}-g^{\mu\sigma}g^{\nu\rho})$}%
}}}}
\put(151,-736){\makebox(0,0)[lb]{\smash{{\SetFigFont{12}{14.4}{\rmdefault}{\mddefault}{\updefault}{\color[rgb]{0,0,0}$Z_{\nu}$}%
}}}}
\put(1276,689){\makebox(0,0)[lb]{\smash{{\SetFigFont{12}{14.4}{\rmdefault}{\mddefault}{\updefault}{\color[rgb]{0,0,0}$W^{n-}_\rho$}%
}}}}
\end{picture}%

%% file: A-W1-W1-Z2.pstex_t
\begin{picture}(0,0)%
\includegraphics{A-W1-W1-Z2.pstex}%
\end{picture}%
\setlength{\unitlength}{3947sp}%
\begingroup\makeatletter\ifx\SetFigFont\undefined%
\gdef\SetFigFont#1#2#3#4#5{%
  \reset@font\fontsize{#1}{#2pt}%
  \fontfamily{#3}\fontseries{#4}\fontshape{#5}%
  \selectfont}%
\fi\endgroup%
\begin{picture}(1725,1639)(151,-794)
\put(151,689){\makebox(0,0)[lb]{\smash{\SetFigFont{12}{14.4}{\rmdefault}{\mddefault}{\updefault}{\color[rgb]{0,0,0}$A_\mu$}%
}}}
\put(1876,-61){\makebox(0,0)[lb]{\smash{\SetFigFont{12}{14.4}{\rmdefault}{\mddefault}{\updefault}{\color[rgb]{0,0,0}$=-i\frac{1}{\sqrt 2} \frac{e^2}{\sin\theta_W} (2g^{\mu\nu}g^{\rho\sigma}-g^{\mu\rho}g^{\nu\sigma}-g^{\mu\sigma}g^{\nu\rho})$}%
}}}
\put(1276,689){\makebox(0,0)[lb]{\smash{\SetFigFont{12}{14.4}{\rmdefault}{\mddefault}{\updefault}{\color[rgb]{0,0,0}$W^{1-}_\rho$}%
}}}
\put(1276,-736){\makebox(0,0)[lb]{\smash{\SetFigFont{12}{14.4}{\rmdefault}{\mddefault}{\updefault}{\color[rgb]{0,0,0}$W^{1+}_\sigma$}%
}}}
\put(151,-736){\makebox(0,0)[lb]{\smash{\SetFigFont{12}{14.4}{\rmdefault}{\mddefault}{\updefault}{\color[rgb]{0,0,0}$Z^2_{\nu}$}%
}}}
\end{picture}

%% file: W-W-W1-W1.pstex_t
\begin{picture}(0,0)%
\includegraphics{W-W-W1-W1.pstex}%
\end{picture}%
\setlength{\unitlength}{3947sp}%
\begingroup\makeatletter\ifx\SetFigFont\undefined%
\gdef\SetFigFont#1#2#3#4#5{%
  \reset@font\fontsize{#1}{#2pt}%
  \fontfamily{#3}\fontseries{#4}\fontshape{#5}%
  \selectfont}%
\fi\endgroup%
\begin{picture}(1755,1657)(136,-809)
\put(1276,-736){\makebox(0,0)[lb]{\smash{{\SetFigFont{12}{14.4}{\rmdefault}{\mddefault}{\updefault}{\color[rgb]{0,0,0}$W^{n-}_{\sigma}$}%
}}}}
\put(151,689){\makebox(0,0)[lb]{\smash{{\SetFigFont{12}{14.4}{\rmdefault}{\mddefault}{\updefault}{\color[rgb]{0,0,0}$W^{+}_\mu$}%
}}}}
\put(151,-736){\makebox(0,0)[lb]{\smash{{\SetFigFont{12}{14.4}{\rmdefault}{\mddefault}{\updefault}{\color[rgb]{0,0,0}$W^+_\nu$}%
}}}}
\put(1876,-61){\makebox(0,0)[lb]{\smash{{\SetFigFont{12}{14.4}{\rmdefault}{\mddefault}{\updefault}{\color[rgb]{0,0,0}$=ig_2^2 (2g^{\mu\nu}g^{\rho\sigma}-g^{\mu\rho}g^{\nu\sigma}-g^{\mu\sigma}g^{\nu\rho})$}%
}}}}
\put(1276,689){\makebox(0,0)[lb]{\smash{{\SetFigFont{12}{14.4}{\rmdefault}{\mddefault}{\updefault}{\color[rgb]{0,0,0}$W^{n-}_\rho$}%
}}}}
\end{picture}%

%% file: W-W-Z1-Z1.pstex_t
\begin{picture}(0,0)%
\includegraphics{W-W-Z1-Z1.pstex}%
\end{picture}%
\setlength{\unitlength}{3947sp}%
\begingroup\makeatletter\ifx\SetFigFont\undefined%
\gdef\SetFigFont#1#2#3#4#5{%
  \reset@font\fontsize{#1}{#2pt}%
  \fontfamily{#3}\fontseries{#4}\fontshape{#5}%
  \selectfont}%
\fi\endgroup%
\begin{picture}(1755,1657)(136,-809)
\put(1276,-736){\makebox(0,0)[lb]{\smash{{\SetFigFont{12}{14.4}{\rmdefault}{\mddefault}{\updefault}{\color[rgb]{0,0,0}$Z^{n}_{\sigma}$}%
}}}}
\put(151,689){\makebox(0,0)[lb]{\smash{{\SetFigFont{12}{14.4}{\rmdefault}{\mddefault}{\updefault}{\color[rgb]{0,0,0}$W^+_\mu$}%
}}}}
\put(151,-736){\makebox(0,0)[lb]{\smash{{\SetFigFont{12}{14.4}{\rmdefault}{\mddefault}{\updefault}{\color[rgb]{0,0,0}$W^-_\nu$}%
}}}}
\put(1876,-61){\makebox(0,0)[lb]{\smash{{\SetFigFont{12}{14.4}{\rmdefault}{\mddefault}{\updefault}{\color[rgb]{0,0,0}$=-ig_2^2 (2g^{\mu\nu}g^{\rho\sigma}-g^{\mu\rho}g^{\nu\sigma}-g^{\mu\sigma}g^{\nu\rho})$}%
}}}}
\put(1276,689){\makebox(0,0)[lb]{\smash{{\SetFigFont{12}{14.4}{\rmdefault}{\mddefault}{\updefault}{\color[rgb]{0,0,0}$Z^{n}_\rho$}%
}}}}
\end{picture}%

%% file: W-W1-Z-Z1.pstex_t
\begin{picture}(0,0)%
\includegraphics{W-W1-Z-Z1.pstex}%
\end{picture}%
\setlength{\unitlength}{3947sp}%
\begingroup\makeatletter\ifx\SetFigFont\undefined%
\gdef\SetFigFont#1#2#3#4#5{%
  \reset@font\fontsize{#1}{#2pt}%
  \fontfamily{#3}\fontseries{#4}\fontshape{#5}%
  \selectfont}%
\fi\endgroup%
\begin{picture}(1755,1657)(136,-809)
\put(1276,-736){\makebox(0,0)[lb]{\smash{{\SetFigFont{12}{14.4}{\rmdefault}{\mddefault}{\updefault}{\color[rgb]{0,0,0}$Z^{n}_{\sigma}$}%
}}}}
\put(151,689){\makebox(0,0)[lb]{\smash{{\SetFigFont{12}{14.4}{\rmdefault}{\mddefault}{\updefault}{\color[rgb]{0,0,0}$W^+_\mu$}%
}}}}
\put(1876,-61){\makebox(0,0)[lb]{\smash{{\SetFigFont{12}{14.4}{\rmdefault}{\mddefault}{\updefault}{\color[rgb]{0,0,0}$=-i\cos\theta_W g_2^2 (2g^{\mu\nu}g^{\rho\sigma}-g^{\mu\rho}g^{\nu\sigma}-g^{\mu\sigma}g^{\nu\rho})$}%
}}}}
\put(1276,689){\makebox(0,0)[lb]{\smash{{\SetFigFont{12}{14.4}{\rmdefault}{\mddefault}{\updefault}{\color[rgb]{0,0,0}$Z_\rho$}%
}}}}
\put(151,-736){\makebox(0,0)[lb]{\smash{{\SetFigFont{12}{14.4}{\rmdefault}{\mddefault}{\updefault}{\color[rgb]{0,0,0}$W^{n-}_\nu$}%
}}}}
\end{picture}%

%% file: W-W1-Z1-Z2.pstex_t
\begin{picture}(0,0)%
\includegraphics{W-W1-Z1-Z2.pstex}%
\end{picture}%
\setlength{\unitlength}{3947sp}%
\begingroup\makeatletter\ifx\SetFigFont\undefined%
\gdef\SetFigFont#1#2#3#4#5{%
  \reset@font\fontsize{#1}{#2pt}%
  \fontfamily{#3}\fontseries{#4}\fontshape{#5}%
  \selectfont}%
\fi\endgroup%
\begin{picture}(1725,1639)(151,-794)
\put(151,689){\makebox(0,0)[lb]{\smash{\SetFigFont{12}{14.4}{\rmdefault}{\mddefault}{\updefault}{\color[rgb]{0,0,0}$W^+_\mu$}%
}}}
\put(151,-736){\makebox(0,0)[lb]{\smash{\SetFigFont{12}{14.4}{\rmdefault}{\mddefault}{\updefault}{\color[rgb]{0,0,0}$W^{1-}_\nu$}%
}}}
\put(1876,-61){\makebox(0,0)[lb]{\smash{\SetFigFont{12}{14.4}{\rmdefault}{\mddefault}{\updefault}{\color[rgb]{0,0,0}$=-i\frac{1}{\sqrt2}\frac{e^2}{\sin\theta_W} (2g^{\mu\nu}g^{\rho\sigma}-g^{\mu\rho}g^{\nu\sigma}-g^{\mu\sigma}g^{\nu\rho})$}%
}}}
\put(1276,689){\makebox(0,0)[lb]{\smash{\SetFigFont{12}{14.4}{\rmdefault}{\mddefault}{\updefault}{\color[rgb]{0,0,0}$Z^{1}_\rho$}%
}}}
\put(1276,-736){\makebox(0,0)[lb]{\smash{\SetFigFont{12}{14.4}{\rmdefault}{\mddefault}{\updefault}{\color[rgb]{0,0,0}$Z^{2}_{\sigma}$}%
}}}
\end{picture}

%% file: W1-W1-W1-W1.pstex_t
\begin{picture}(0,0)%
\includegraphics{W1-W1-W1-W1.pstex}%
\end{picture}%
\setlength{\unitlength}{3947sp}%
\begingroup\makeatletter\ifx\SetFigFont\undefined%
\gdef\SetFigFont#1#2#3#4#5{%
  \reset@font\fontsize{#1}{#2pt}%
  \fontfamily{#3}\fontseries{#4}\fontshape{#5}%
  \selectfont}%
\fi\endgroup%
\begin{picture}(1755,1657)(136,-809)
\put(1276,-736){\makebox(0,0)[lb]{\smash{{\SetFigFont{12}{14.4}{\rmdefault}{\mddefault}{\updefault}{\color[rgb]{0,0,0}$W^{n-}_{\sigma}$}%
}}}}
\put(1876,-61){\makebox(0,0)[lb]{\smash{{\SetFigFont{12}{14.4}{\rmdefault}{\mddefault}{\updefault}{\color[rgb]{0,0,0}$=i\frac{3}{2}g_2^2 (2g^{\mu\nu}g^{\rho\sigma}-g^{\mu\rho}g^{\nu\sigma}-g^{\mu\sigma}g^{\nu\rho})$}%
}}}}
\put(151,689){\makebox(0,0)[lb]{\smash{{\SetFigFont{12}{14.4}{\rmdefault}{\mddefault}{\updefault}{\color[rgb]{0,0,0}$W^{n+}_\mu$}%
}}}}
\put(1276,689){\makebox(0,0)[lb]{\smash{{\SetFigFont{12}{14.4}{\rmdefault}{\mddefault}{\updefault}{\color[rgb]{0,0,0}$W^{n-}_\rho$}%
}}}}
\put(151,-736){\makebox(0,0)[lb]{\smash{{\SetFigFont{12}{14.4}{\rmdefault}{\mddefault}{\updefault}{\color[rgb]{0,0,0}$W^{n+}_\nu$}%
}}}}
\end{picture}%

%% file: W1-W1-Z-Z.pstex_t
\begin{picture}(0,0)%
\includegraphics{W1-W1-Z-Z.pstex}%
\end{picture}%
\setlength{\unitlength}{3947sp}%
\begingroup\makeatletter\ifx\SetFigFont\undefined%
\gdef\SetFigFont#1#2#3#4#5{%
  \reset@font\fontsize{#1}{#2pt}%
  \fontfamily{#3}\fontseries{#4}\fontshape{#5}%
  \selectfont}%
\fi\endgroup%
\begin{picture}(1755,1657)(136,-809)
\put(151,689){\makebox(0,0)[lb]{\smash{{\SetFigFont{12}{14.4}{\rmdefault}{\mddefault}{\updefault}{\color[rgb]{0,0,0}$W^{n+}_\mu$}%
}}}}
\put(1876,-61){\makebox(0,0)[lb]{\smash{{\SetFigFont{12}{14.4}{\rmdefault}{\mddefault}{\updefault}{\color[rgb]{0,0,0}$=-i\cos^2 \theta_W g_2^2 (2g^{\mu\nu}g^{\rho\sigma}-g^{\mu\rho}g^{\nu\sigma}-g^{\mu\sigma}g^{\nu\rho})$}%
}}}}
\put(1276,689){\makebox(0,0)[lb]{\smash{{\SetFigFont{12}{14.4}{\rmdefault}{\mddefault}{\updefault}{\color[rgb]{0,0,0}$Z_\rho$}%
}}}}
\put(1276,-736){\makebox(0,0)[lb]{\smash{{\SetFigFont{12}{14.4}{\rmdefault}{\mddefault}{\updefault}{\color[rgb]{0,0,0}$Z_{\sigma}$}%
}}}}
\put(151,-736){\makebox(0,0)[lb]{\smash{{\SetFigFont{12}{14.4}{\rmdefault}{\mddefault}{\updefault}{\color[rgb]{0,0,0}$W^{n-}_\nu$}%
}}}}
\end{picture}%

%% file: W1-W1-Z-Z2.pstex_t
\begin{picture}(0,0)%
\includegraphics{W1-W1-Z-Z2.pstex}%
\end{picture}%
\setlength{\unitlength}{3947sp}%
\begingroup\makeatletter\ifx\SetFigFont\undefined%
\gdef\SetFigFont#1#2#3#4#5{%
  \reset@font\fontsize{#1}{#2pt}%
  \fontfamily{#3}\fontseries{#4}\fontshape{#5}%
  \selectfont}%
\fi\endgroup%
\begin{picture}(1725,1639)(151,-794)
\put(151,689){\makebox(0,0)[lb]{\smash{\SetFigFont{12}{14.4}{\rmdefault}{\mddefault}{\updefault}{\color[rgb]{0,0,0}$W^{1+}_\mu$}%
}}}
\put(151,-736){\makebox(0,0)[lb]{\smash{\SetFigFont{12}{14.4}{\rmdefault}{\mddefault}{\updefault}{\color[rgb]{0,0,0}$W^{1-}_\nu$}%
}}}
\put(1876,-61){\makebox(0,0)[lb]{\smash{\SetFigFont{12}{14.4}{\rmdefault}{\mddefault}{\updefault}{\color[rgb]{0,0,0}$=-i\frac{1}{\sqrt2}\cos\theta_W g_2^2 (2g^{\mu\nu}g^{\rho\sigma}-g^{\mu\rho}g^{\nu\sigma}-g^{\mu\sigma}g^{\nu\rho})$}%
}}}
\put(1276,689){\makebox(0,0)[lb]{\smash{\SetFigFont{12}{14.4}{\rmdefault}{\mddefault}{\updefault}{\color[rgb]{0,0,0}$Z_\rho$}%
}}}
\put(1276,-736){\makebox(0,0)[lb]{\smash{\SetFigFont{12}{14.4}{\rmdefault}{\mddefault}{\updefault}{\color[rgb]{0,0,0}$Z^{2}_{\sigma}$}%
}}}
\end{picture}

%% file: W1-W1-Z1-Z1.pstex_t
\begin{picture}(0,0)%
\includegraphics{W1-W1-Z1-Z1.pstex}%
\end{picture}%
\setlength{\unitlength}{3947sp}%
\begingroup\makeatletter\ifx\SetFigFont\undefined%
\gdef\SetFigFont#1#2#3#4#5{%
  \reset@font\fontsize{#1}{#2pt}%
  \fontfamily{#3}\fontseries{#4}\fontshape{#5}%
  \selectfont}%
\fi\endgroup%
\begin{picture}(1755,1657)(136,-809)
\put(1276,-736){\makebox(0,0)[lb]{\smash{{\SetFigFont{12}{14.4}{\rmdefault}{\mddefault}{\updefault}{\color[rgb]{0,0,0}$Z^{n}_{\sigma}$}%
}}}}
\put(1876,-61){\makebox(0,0)[lb]{\smash{{\SetFigFont{12}{14.4}{\rmdefault}{\mddefault}{\updefault}{\color[rgb]{0,0,0}$=-i\frac{3}{2}g_2^2 (2g^{\mu\nu}g^{\rho\sigma}-g^{\mu\rho}g^{\nu\sigma}-g^{\mu\sigma}g^{\nu\rho})$}%
}}}}
\put(151,689){\makebox(0,0)[lb]{\smash{{\SetFigFont{12}{14.4}{\rmdefault}{\mddefault}{\updefault}{\color[rgb]{0,0,0}$W^{n+}_\mu$}%
}}}}
\put(1276,689){\makebox(0,0)[lb]{\smash{{\SetFigFont{12}{14.4}{\rmdefault}{\mddefault}{\updefault}{\color[rgb]{0,0,0}$Z^{n}_\rho$}%
}}}}
\put(151,-736){\makebox(0,0)[lb]{\smash{{\SetFigFont{12}{14.4}{\rmdefault}{\mddefault}{\updefault}{\color[rgb]{0,0,0}$W^{n-}_\nu$}%
}}}}
\end{picture}%

%% file: W1-W1-Z2-Z2.pstex_t
\begin{picture}(0,0)%
\includegraphics{W1-W1-Z2-Z2.pstex}%
\end{picture}%
\setlength{\unitlength}{3947sp}%
\begingroup\makeatletter\ifx\SetFigFont\undefined%
\gdef\SetFigFont#1#2#3#4#5{%
  \reset@font\fontsize{#1}{#2pt}%
  \fontfamily{#3}\fontseries{#4}\fontshape{#5}%
  \selectfont}%
\fi\endgroup%
\begin{picture}(1725,1639)(151,-794)
\put(151,689){\makebox(0,0)[lb]{\smash{\SetFigFont{12}{14.4}{\rmdefault}{\mddefault}{\updefault}{\color[rgb]{0,0,0}$W^{1+}_\mu$}%
}}}
\put(151,-736){\makebox(0,0)[lb]{\smash{\SetFigFont{12}{14.4}{\rmdefault}{\mddefault}{\updefault}{\color[rgb]{0,0,0}$W^{1-}_\nu$}%
}}}
\put(1876,-61){\makebox(0,0)[lb]{\smash{\SetFigFont{12}{14.4}{\rmdefault}{\mddefault}{\updefault}{\color[rgb]{0,0,0}$=-i\frac{1}{2}g_2^2 (2g^{\mu\nu}g^{\rho\sigma}-g^{\mu\rho}g^{\nu\sigma}-g^{\mu\sigma}g^{\nu\rho})$}%
}}}
\put(1276,689){\makebox(0,0)[lb]{\smash{\SetFigFont{12}{14.4}{\rmdefault}{\mddefault}{\updefault}{\color[rgb]{0,0,0}$Z^{2}_\rho$}%
}}}
\put(1276,-736){\makebox(0,0)[lb]{\smash{\SetFigFont{12}{14.4}{\rmdefault}{\mddefault}{\updefault}{\color[rgb]{0,0,0}$Z^{2}_{\sigma}$}%
}}}
\end{picture}